\documentclass{article}
\usepackage{amssymb}
\usepackage{amsmath}
\usepackage{epsfig}

\let\tsection\section
\renewcommand{\section}{\setcounter{equation}{0}\tsection}

\begin{document}

\begin{titlepage}

~\vspace{-1.7cm}

\hspace{-0.5cm}{\bf The Asymmetric Avalanche Process}

\vspace{1cm}

\begin{center}

{ \LARGE \bf The Asymmetric Avalanche Process.}

\bigskip

{\large A.M. Povolotsky \footnote{Institute of Physics, Academia Sinica, Nankang, Taipei 11529, Taiwan,  huck@phys.sinica.edu.tw}$^,$
\footnote{Bogoliubov Laboratory of \ Theoretical Physics, J.I.N.R., Dubna 141980, Russia, povam@thsun1.jinr.ru, priezzvb@thsun1.jinr.ru},
V.B. Priezzhev $^2$ ,
Chin-Kun Hu $^1$}
\end{center}

\bigskip

\begin{abstract}
An asymmetric stochastic process describing the avalanche dynamics on a ring
is proposed. A general kinetic equation which incorporates the exclusion and
avalanche processes is considered. The Bethe ansatz method is used to calculate
the generating function for the total distance covered by all
\ particles. It gives the average velocity of particles which exhibits a
phase transition from an intermittent to continuous flow. We calculated also
higher cumulants and the large deviation function for the particle flow. The
latter has the universal form obtained earlier for the asymmetric exclusion
process and conjectured to be common for all models of the
Kardar-Parisi-Zhang universality class .
\end{abstract}


%\vspace{2cm}

%Contact information for corresponding author:
%
%\noindent
%A.M. Povolotsky, Institute of Physics, Academia Sinica,
%
%\noindent
%Nankang,Taipei, 11529 Taiwan, e-mail: povam@phys.sinica.edu.tw
%
%\noindent
%Fax: +886-2-27834187, ~~~~ Telephone: +886 2-2789-6762
%
%

\end{titlepage}

\section{Introduction\protect\bigskip .}

Interacting particle systems with stochastic dynamics \cite{Ligget} and
particularly the one-dimensional asymmetric exclusion process (ASEP) have
been intensively studied \cite{Privman}-\cite{pmb}, due to connections to
growth processes \cite{hz}, traffic flows \cite{shrek}, the noisy Burgers
equation \cite{fns} and the Kardar-Parisi-Zhang (KPZ) equation \cite{KPZ}.
Being one of the simplest examples of integrable non-equilibrium systems,
the ASEP allows calculation of some dynamical properties \cite{dhar},\cite%
{gs}, a large deviation function \cite{dl}, and conditional probabilities
\cite{Schutz}.

In a standard formulation \cite{derrida}, particles move in such a way that
there is always at most one particle per site on the one-dimensional
lattice. Every particle hops to its right or left with biased probabilities
provided the target site is empty. Otherwise, it does not move. Using a
traffic terminology, this kind of interaction between particles can be
called a ``soft braking''.

Another kind of interaction is an \textquotedblleft aggressive
braking\textquotedblright\ \cite{srb},\cite{akk2}, when a particle pushes
the particle in front of it and then stops. The displaced particle shifts
the next particle in front, if any, and so on. As a result, a chain of
adjacent particles is shifted by one lattice space left or right at the same
moment of time. Despite apparent non-locality of dynamics, the Bethe ansatz
method is still applicable and the resulting Bethe equations are solvable
parallel to the ASEP case. Further generalizations of the ASEP have been
proposed \cite{sw},\cite{khor}. In every case, however, an elementary motion
of a particle produces a deterministic reconstruction (local or non-local)
of the preceding lattice state.

The beginning of intensive study of the ASEP nearly coincides with a burst
of interest to the threshold dynamics and avalanche processes. Appeared
originally in the sandpile model of self-organized criticality \cite{btw},
the avalanche processes have been shown to be related to many different
phenomena ranging from an interface depinning to earthquakes \cite{bak}.

As an example of the threshold dynamics one can again consider ASEP-like
stochastic model at the one dimensional \ lattice. In this case, however, we
admit multiple occupation of a lattice site by particles. Like in the ASEP,
each particle hops to its right or left. If the number of particles $n$ at
given site exceeds some critical value $n_{c}$, the site is unstable and
must relax immediately. The relaxation consists in spilling of $%
m\eqslantless n$ particles from the given site to neighboring sites by a
fixed rule. If the neighboring sites become unstable, they relax as well.
Thus, an avalanche of relaxations spreads over the lattice. The time
interval between beginning and ending of every avalanche is negligible in
comparison with characteristic hopping time of a single particle.

Comparing to the ASEP with the aggressive braking, a fundamental difference
appears, when the spilling rule is stochastic \cite{manna}. In this case,
the structure of avalanche becomes complicated. Unstable states may appear
randomly even if an underlying structure of the lattice state is regular
before an avalanche starts. Then, the distance at which avalanches propagate
and the total mass of particles involved in an avalanche are random values
described by probabilistic distributions. The configurations of particles in
the lattice states before and after an avalanche may differ considerably,
and the latter results from the first by a series of stochastic spillings.

Another peculiarity of avalanche dynamics is a specific transition into a
totally unstable state, when the density of particles exceeds some critical
value and an avalanche never stops in the thermodynamic limit of infinitely
large lattice \cite{Vespig}. This transition corresponds to change of the
time scale characterizing the system, which can be defined for example as a
ratio of system size to average velocity of particles. While for low
density, the slow diffusion processes prevail, the fast avalanches dominate
above the transition point. The existing of two time scales was shown to be
responsible for reaching of the self-organized critical state in systems
with avalanche dynamics.

The aim of this paper is to give a mathematical description of one kind of
avalanche processes, the one-dimensional asymmetric avalanche process (ASAP)%
\cite{piph}. The ASAP is a partially asymmetric diffusion process with the
totally asymmetric avalanche propagation. The similar directed
one-dimensional stochastic avalanches have been considered in \cite{MZ},
where the asymptotic of avalanche distributions in the self-organized
critical state have been calculated exactly for the open lattice in the
thermodynamic limit. Instead, we study the ASAP on a ring with a fixed
number of particles. In this case, the critical value of the density exists,
depending on spilling probabilities, which corresponds to the transition
from the intermittent to continuous flow when the fast avalanche dynamics
becomes dominating. In this paper, we concentrate on dynamical properties of
the ASAP below this point.

One of the reasons for the intensive interest to the ASEP is that it being
exactly solvable gives a discrete version of the Kardar-Parisi-Zhang (KPZ)
equation \cite{gs}. In the scaling limit, one can get analytically the
universal quantities like critical exponents and scaling functions
characterizing a vast class of nonequilibrium phenomena belonging to the KPZ
universality class. On the other hand, the universal scaling properties of
avalanche dynamics are much less investigated. There are very few successful
attempts, \cite{JKK}-\cite{cn}, to find analytical arguments allowing one to
relate the avalanche-like processes with one of well-defined universality
classes such as the KPZ , Edwards-Wilkinson or directed percolation \cite{O}.

In the present work, we show that the generating function of the total
distance $Y_{t}$ travelled by particles in the ASAP is given in the scaling
limit by the expression
\begin{equation}
\lim\limits_{t\rightarrow \infty }\frac{\ln \left\langle e^{\gamma
Y_{t}}\right\rangle }{t}\sim \gamma K_{1}+K_{2}G(K_{3}\gamma ),
\label{Generating function}
\end{equation}%
where $G(x)$ does not depend on parameters of the model and has the
following parametric form
\begin{eqnarray}
G(x) &=&-\sum_{s=1}^{\infty }(-C)^{s}s^{-5/2}  \label{G(c)} \\
x &=&\sum_{s=1}^{\infty }(-C)^{s}s^{-3/2},  \label{x(c)}
\end{eqnarray}%
and $K_{1},K_{2},$ and $K_{3}$ are model dependent parameters. This
universal form of function $G(x)$ was claimed to be the feature of the KPZ
universality class \cite{da},\cite{DB}. It also determines the universal
form of the large deviation function characterizing the deviations of the
integrated particle current from its average value. These results give an
evidence that the ASAP\ shares the KPZ universality class with the ASEP
despite the significant difference in their dynamics.

The article is organized as follows. In Section 2, we consider the master
equation for a general stochastic model which leads to the ASEP and ASAP in
particular cases. In Section 3, we derive the Bethe ansatz equations for the
generating function of the total displacement of all particles. A particular
case of the ASAP is considered in Section 4 where the ASAP becomes identical
to a drop-push version of the ASEP. The general case of the ASAP is
considered in Section 5. Using the method developed in \cite{k},\cite{LK},
we analyse the integral equation corresponding to the density of roots of
the Bethe equations and derive the generating function for the total
displacement of particles. From it we obtain the expression for the
cumulants of total distance travelled by particles, like mean velocity and
variance, its large deviation function, and demonstrate that the ASAP
belongs to the KPZ universality class.\footnote{A brief summary of the results has been presented at
"StatPhys-Taiwan 2002", during May 27 -June 1, see \cite{pph1}}
\section{The master equation.}

In this section we are going to obtain the master equation describing ASAP,
which is defined as follows. Consider the system of $p$ particles on a ring
of $\ N$ sites as shown in Fig. \ref{ASAP}.
Particles jump left or right with probabilities $Ldt$ or $Rdt,$
respectively, for infinitesimal time $dt$ independently of each other. When
a particle comes to already occupied site after the hopping either from left
with the rate $R$ or from right with the rate $L$ , an avalanche starts. It
develops step by step according to the following dynamical rules.

If, at some step of the avalanche, $n$ $\ (n=2,3,...)$  particles are at
site $x$, then

with probability $\mu _{n}$ , $n$\ particles go to the site $x+1$ ;

with probability $1-\mu _{n}$, $n-1$ particles go to the site $x+1$ and one
particle stays at the current site $x$.

We imply that an avalanche takes infinitesimal time to end, i.e. from the
point of view of Poissonian processes it plays a role of interaction
resulting in the transition between configurations with single particle
occupation. The totally asymmetric case discussed in \cite{piph}
corresponds to particular choice of the rates, $L=0,R=1$. In the case of the
ASEP a particle step to already occupied site is forbidden. However, it will
be shown below to be closely connected with the ASAP. To make the
presentation more systematic we start from the system of free particles then
going to ASEP and ASAP dynamics.

The state $C$ of the system at time $t$ is characterized by the probability $%
P_{t}(C)$ satisfying the master equation
\begin{equation}
\partial _{t}P_{t}(C)=\sum_{\{C^{\prime }\}}M(C,C^{\prime })P_{t}(C^{\prime
})  \label{mastereq}
\end{equation}%
The off-diagonal elements $M(C,C^{\prime })$ of the matrix $M$ are rates of
transitions from configurations $C^{\prime }$ to $C$ and therefore are
always positive. The diagonal elements $M(C,C)$ that give the total rate of
the transition from the state $C$ to all other configurations, enter the
matrix $M$ with a minus sign. Conservation of probability requires the
identity
\begin{equation}
M(C,C)=-\sum_{\{C^{\prime }\}}M(C^{\prime },C)  \label{conserv}
\end{equation}

Let us return to particles at the lattice. Consider noninteracting particles
jumping left or right with probabilities $Ldt$ or $Rdt,$ respectively, for
infinitesimal time $dt$ . The probability $P_{t}(x_{1},\ldots ,x_{p})$ for
particles to occupy sites $x_{1},\ldots ,x_{p}$ obeys the master equation
\begin{eqnarray}
&&\partial _{t}P_{t}(x_{1},\ldots ,x_{p})=-pP_{t}(x_{1},\ldots ,x_{p})+
\label{master1} \\
&&L\sum_{i=1}^{p}P_{t}(x_{1},\ldots ,x_{i}+1,\ldots
,x_{p})+R\sum_{i=1}^{p}P_{t}(x_{1},\ldots ,x_{i}-1,\ldots ,x_{p})  \notag
\end{eqnarray}%
if $x_{i+1}-x_{i}>1$.We impose the condition $L+R=1$ by an appropriate
choice of time scale.

In the ASEP, the form of the master equation should be modified if
configuration $C$ contains pairs of neighboring occupied sites. If there are
only two neighboring sites, $x,x+1,$ occupied by particles,\ the master
equation differs from Eq.(\ref{master1}) by the extra term $LP_{t}(\ldots
,x+1,x+1,\ldots )+RP_{t}(\ldots ,x,x,\ldots )-P_{t}(\ldots ,x,x+1,\ldots )$.
If \ there are more than one pair, one must substract the other unwanted
terms from the Eq.(\ref{master1}) for every pair to obtain equations taking
into account the exclusion rules. Instead, one can reduce the consideration
to the free master equation only, if one equates the appeared extra terms to
zero\ putting the boundary conditions for the physical domain $%
x_{1}<x_{2}<...<x_{p}$ :

\begin{equation}
LP_{t}(\ldots ,x+1,x+1,\ldots )+RP_{t}(\ldots ,x,x,\ldots )-P_{t}(\ldots
,x,x+1,\ldots )=0.  \label{asepBC}
\end{equation}
The terms like $P_{t}(\ldots ,x,x,\ldots )$ corresponding to multiple
occupation of sites do not contribute to the dynamics due to the exclusion
rule. Therefore, they can be considered as auxiliary non-physical terms and
redefined by the boundary conditions so that all extra terms in free
equation vanish giving the correct equation for the system with interaction.

Consider now more general condition of type Eq.(\ref{asepBC}) with the
coefficients $\alpha $ and $\beta $ which do not coincide necessarily with
the rates $L$ and $R$ in Eq.(\ref{master1}) and will be defined later,
\begin{equation}
\alpha P_{t}(\ldots ,x,x,\ldots )+\beta P_{t}(\ldots ,x+1,x+1,\ldots
)-P_{t}(\ldots ,x,x+1,\ldots )=0  \label{ASAPBCgeneral}
\end{equation}

To provide the probabilistic meaning of the Eq.(\ref{master1}) together with
Eq.(\ref{ASAPBCgeneral}), some constraints should be imposed on $\alpha $
and $\beta $ . In this case, we still use the exclusion rule that allows one
to use the terms of type $P_{t}(\ldots ,x,x,\ldots )$ as auxiliary ones
which should be redefined in appropriate way. The condition Eq.(\ref%
{ASAPBCgeneral}) itself does not eliminate the contribution of extra terms
yet. Nevertheless, we can try to use this condition \ to replace the
unwanted terms by terms consisting of allowed configurations only. To this
end, we can exploit the fact that two unphysical terms in Eq.(\ref%
{ASAPBCgeneral}) are of similar structure and consider this relation as a
recursion:
\begin{equation}
P_{t}(\ldots ,x,x,\ldots )=\frac{1}{\alpha }P_{t}(\ldots ,x,x+1,\ldots )-%
\frac{\beta }{\alpha }P_{t}(\ldots ,x+1,x+1,\ldots ),  \label{x,x->x+1,x+1}
\end{equation}%
To proceed, it is convenient to consider the two-particle case separately.

\subsection{The case of two particles.}

If there are only two particles at the lattice, the recursion Eq.(\ref%
{x,x->x+1,x+1}) can be immediately solved in terms of allowed configurations
only:

\begin{equation}
P_{t}(x,x)=\frac{1}{\beta }\sum_{n=0}^{\infty }\left( -\frac{\alpha }{\beta }%
\right) ^{n}P_{t}(x-n-1,x-n)  \label{rec2left}
\end{equation}%
where, due to periodic boundary conditions, all coordinates are integers
modulo $N$ . Substituting Eq.(\ref{rec2left}) into the Eq.(\ref{master1})
rewritten for the two particle case under the condition $x_{2}=x_{1}+1$, we
get%
\begin{eqnarray}
&&\!\!\!\!\!\!\!\!\partial
_{t}P_{t}(x,x+1)=LP_{t}(x,x+2)+RP_{t}(x-1,x+1)-P_{t}(x,x+1)  \label{master2}
\\
&&\!\!\!\!\!\!\!+\left( \frac{L}{\beta }-1\right) P_{t}(x,x+1)+\frac{L}{%
\beta }(\mu +\frac{R}{L})\sum_{n=1}^{\infty }\mu ^{n-1}P_{t}(x-n,x-n+1)
\notag
\end{eqnarray}%
where $\mu =-\alpha /\beta $. \ To give a probabilistic meaning to the
transition rates, the terms corresponding to the processes when the system
leaves the configuration $(x,x+1)$ should be non-positive and those for
coming into $(x,x+1)$ from other configurations should be non-negative. To
keep probabilities positive, we have to impose the condition that either $%
\mu $ is positive or $\mu =-R/L$,\ when the term containing the infinite sum
vanishes. In addition, conservation of probability, Eq.(\ref{conserv}),
requires
\begin{equation}
\alpha +\beta =1.  \label{conv1}
\end{equation}

Then, the condition $\mu =-R/L$ \ implies $\alpha =R$, $\beta =L$, i.e. the
ordinary ASEP. In the case $\mu >0$ , we have
\begin{equation}
\alpha =-\mu /(1-\mu ),\beta =1/(1-\mu )
\end{equation}%
and can rewrite Eq.(\ref{master2}) in the form
\begin{eqnarray}
&&\!\!\!\!\!\!\!\!\!\!\!\!\!\!\!\!\!\!\!\!\!\partial
_{t}P_{t}(x,x+1)=LP_{t}(x,x+2)+RP_{t}(x-1,x+1)-P_{t}(x,x+1)  \label{ex2} \\
&&\!\!\!\!\!\!\!\!\!\!\!\!\!\!\!\!\!\!\!\!\!+(R+L\mu )(-P_{t}(x,x+1)+(1-\mu
)\sum_{n=1}^{\infty }\mu ^{n-1}P_{t}(x-n,x-n+1)  \notag
\end{eqnarray}%
In terms of $\mu $ , the boundary condition, Eq.(\ref{x,x->x+1,x+1}) reads
\begin{equation}
P_{t}(x,x)=\left( 1-\mu \right) P_{t}(x-1,x)+\mu P_{t}(x-1,x-1)
\label{bouncond2}
\end{equation}%
The expression Eq.(\ref{ex2}) shows that in addition to the Poissonian
hopping given by the original kinetic equation, new terms appear in the
equation which correspond to transitions to the configuration $C=(x,x+1)$
from the configurations $\{C^{\prime }\}=\{(x-n,x-n+1),$ $n=1,2,...\}$. The
rates of the transitions are $(R+L\mu )(1-\mu )\mu ^{n-1}$. In the case of
two particles,these rates determine the avalanche dynamics defined above.

To show the relation between ASEP and ASAP before going into details of the
solution, let us consider the moment when an avalanche starts. This happens
if either the left particle of two neighboring ones moves right with the
rate $R$ or the right particle moves to the left with the rate $L$ and then
\ two particles together make at least one step together with probability $%
\mu $. Then an avalanche starts with the rate $(R+L\mu )$. This expression
indeed enters those parts of kinetic equation, which correspond to the
avalanche dynamics. Thus, the rate of beginning of an avalanche becomes zero
when $\mu =-R/L$, and only exclusion dynamics remains. Therefore, we may
treat ASEP as an analytical continuation of ASAP with a parameter $\mu $
taking a special negative value.

\subsection{Many particle processes.}

\ One can expect that the $n$-particle interactions imposes $n$ new
constrains on the master equation Eq.(\ref{master1}). However, in this
section we will show that under certain constraint on toppling probabilities
$\mu _{n}$ no new boundary conditions appear and Eq.(\ref{bouncond2}) is
sufficient to take into account the interaction of arbitrary number of
particles.

To generalize the boundary condition, Eq.(\ref{bouncond2}), for the
description of many particle dynamics defined above one should express an
unstable configuration via configurations appeared at the previous steps of
an avalanche. The form of these conditions depends on the fact whether the
site $x-1$ is occupied or not:

\begin{eqnarray}
&&P_{t}(\ldots ,x-1,\underset{n-1}{\underbrace{x,\ldots ,x}},\ldots )=(1-\mu
_{n})P_{t}(\ldots ,\underset{n}{\underbrace{x-1,\ldots ,x-1}},\ldots )+
\notag \\
&&(1-\mu _{n-1})P_{t}(\ldots ,\underset{n-1}{\underbrace{x-1,\ldots ,x-1}}%
,x,\ldots )  \label{boundcond3}
\end{eqnarray}%
if the site $x-1$ is occupied, and
\begin{eqnarray}
&&P_{t}(\ldots ,\underset{n}{\underbrace{x,\ldots ,x}},\ldots )=\mu
_{n}P_{t}(\ldots ,\underset{n}{\underbrace{x-1,\ldots ,x-1}},\ldots )+
\notag \\
&&\mu _{n-1}P_{t}(\ldots ,\underset{n-1}{\underbrace{x-1,\ldots ,x-1}}%
,x,\ldots )  \label{boundcond4}
\end{eqnarray}%
if $x-1$ is empty.

Like the two-particle boundary condition, Eq.(\ref{bouncond2}), the
many-particle conditions, Eqs.(\ref{boundcond3}) and (\ref{boundcond4}),
should be applied recursively. Applying this recursion step by step to
infinity, we generate an infinite series that consist of the transition
probabilities between stable configurations only.

On the other hand, one may treat the term $P_{t}(\ldots ,\underset{n}{%
\underbrace{x,\ldots ,x},}\ldots )$ formally applying the two-particle
boundary condition Eq.(\ref{bouncond2}), reducing sequentially the number of
particles in unstable sites. As a result, we obtain

\begin{eqnarray}
&&P_{t}(\ldots ,\underset{n}{\underbrace{x,\ldots ,x}},\ldots )=\mu
_{n}P_{t}(\ldots ,\underset{n}{\underbrace{x-1,\ldots ,x-1}},\ldots )+
\notag \\
&&(1-\mu _{n})P_{t}(\ldots ,\underset{n-1}{\underbrace{x-1,\ldots ,x-1}}%
,x,\ldots )  \label{boundcond5}
\end{eqnarray}
\ Due to recursion, parameters $\mu _{n}$ are expressed through the only
parameter $\mu $:
\begin{equation}
\mu _{2}=\mu ,\quad \mu _{3}=\left( 1-\mu \right) \mu ,\quad \mu _{n}=\left(
1-\mu \right) \mu _{n-1}+\mu \mu _{n-2},\quad n>3  \label{munrec}
\end{equation}
or
\begin{equation}
\mu _{n}=\mu \frac{1-(-\mu )^{n-1}}{1+\mu }.  \label{mun}
\end{equation}
Generally, Eq.(\ref{boundcond5}) and Eqs.(\ref{boundcond3}),(\ref{boundcond4}%
) do not coincide. However, the final series entering the kinetic equation
which result from the sequential use of either latter or two former of them
will be the same, provided that the transition probabilities $\mu _{n}$ from
Eqs.(\ref{boundcond3}) and (\ref{boundcond4}) governing the avalanche
dynamics satisfy Eqs.(\ref{mun}) and (\ref{munrec}).

Indeed, every term of resulting series represents a finite avalanche. It
contains the product of terms $\mu _{n}$ and $(1-\mu _{n})$ coming from the
successive use of the recurrent relations, Eqs.(\ref{boundcond3}),(\ref%
{boundcond4}) or Eq.(\ref{boundcond5}). The first term of the r.h.s of Eq.(%
\ref{boundcond3}) increases the number of particles at the unstable site by
1 in comparison with the l.h.s. \ The second term of the r.h.s \ of \ Eq.(%
\ref{boundcond4}) decreases the number of particles by 1. Obviously, in
every finite avalanche, the numbers of decreasing and increasing events are
equal. Therefore, the coefficients $(1-\mu _{n})$ from Eq.(\ref{boundcond3})
and $\mu _{n-1}$ from Eq.(\ref{boundcond4}) always enter the product
corresponding to the avalanche in pairs and interchanging their places do
not affect the structure of the final series. On the other hand, the
interchanging $\mu _{n-1}$ and $(1-\mu _{n})$ between Eqs.(\ref{boundcond3})
and (\ref{boundcond4}) leads to Eq.(\ref{boundcond5}) obtained from Eq.(\ref%
{bouncond2}).

Thus, we have shown that the two-particle bo\-un\-da\-ry con\-di\-ti\-on, Eq.(\ref{bouncond2}),
is sufficient for obtaining the kinetic equation for many
particle avalanche process governed by the above dynamical rules. In the
Bethe ansatz formalism, a similar procedure is known as the two- particle
reducibility \ \cite{bik} and provides integrability of a system.

\section{The Bethe Ansatz for generating function.}

Consider the generating function of the total path $Y$ travelled by all
particles between time $0$ and $t$ provided that the system is in
configuration $C$ at time $t$
\begin{equation}
F_{t}(C)=\sum_{Y=-\infty }^{\infty }P_{t}(C,Y)e^{\gamma Y},
\end{equation}
where $P_{t}(C,Y)$ is the joint probability that the system is in
configuration $C$ and the total distance is $Y$ to the time $t.$ The
equation for the generating function and a boundary condition can be
obtained from the master equation Eq.(\ref{master1}) and the boundary
condition for the probability Eq.(\ref{bouncond2}) by multiplying by $%
e^{\gamma }(e^{-\gamma })$ every term corresponding to increasing
(decreasing) distance $Y$ \ by $1$.

\begin{eqnarray}
&&\partial _{t}F_{t}(x_{1},\ldots ,x_{p})=Le^{-\gamma
}\sum_{i=1}^{p}F_{t}(x_{1},\ldots ,x_{i}+1,\ldots ,x_{p})+  \label{masterF}
\\
&&+Re^{\gamma }\sum_{i=1}^{p}F_{t}(x_{1},\ldots ,x_{i}-1,\ldots
,x_{p})-pF_{t}(x_{1},\ldots ,x_{p})  \notag
\end{eqnarray}
\begin{eqnarray}
F_{t}(\ldots ,x,x,\ldots ) &=&(1-\mu )e^{\gamma }F_{t}(\ldots ,x-1,x,\ldots
)+  \label{boundaryF} \\
&&\mu e^{2\gamma }F_{t}(\ldots ,x-1,x-1,\ldots ).  \notag
\end{eqnarray}

The sum of the function $F_{t}(C)$ over all configurations gives the
generating function of moments of total distance $Y_{t}$, whose behavior is
determined by the largest eigenvalue $\Lambda (\gamma )$ of Eq.(\ref{masterF}%
) for large $t$ .
\begin{equation}
\sum_{\{C\}}F_{t}(C)=\left\langle e^{\gamma Y_{t}}\right\rangle \thicksim
e^{\Lambda (\gamma )t}.
\end{equation}
The derivatives of $\Lambda (\gamma )$ at $\gamma =0$ give the cumulants of
the distance $Y_{t}$
\begin{align}
\lim\limits_{t\rightarrow \infty }\frac{\left\langle Y_{t}\right\rangle _{c}%
}{t}& =\lim\limits_{t\rightarrow \infty }\frac{\left\langle
Y_{t}\right\rangle }{t}=\left. \frac{\partial \Lambda (\gamma )}{\partial
\gamma }\right| _{\gamma =0}  \label{Y1} \\
\lim\limits_{t\rightarrow \infty }\frac{\left\langle Y_{t}^{2}\right\rangle
_{c}}{t}& =\lim\limits_{t\rightarrow \infty }\frac{\left\langle
Y_{t}^{2}\right\rangle -\left\langle Y_{t}\right\rangle ^{2}}{t}=\left.
\frac{\partial ^{2}\Lambda (\gamma )}{\partial \gamma ^{2}}\right| _{\gamma
=0}  \label{Y2} \\
\bigskip \lim\limits_{t\rightarrow \infty }\frac{\left\langle
Y_{t}^{3}\right\rangle _{c}}{t}& =\lim\limits_{t\rightarrow \infty }\frac{%
\left\langle Y_{t}^{3}\right\rangle +2\left\langle Y_{t}\right\rangle
^{3}-3\left\langle Y_{t}^{2}\right\rangle \left\langle Y_{t}\right\rangle }{t%
}=\left. \frac{\partial ^{3}\Lambda (\gamma )}{\partial \gamma ^{3}}\right|
_{\gamma =0}  \label{Y3}
\end{align}

The quantity of our main interest is the large deviation function
\begin{equation}
f(y)=\lim\limits_{t\rightarrow \infty }\frac{1}{t}\ln \mathrm{Prob}\left(
\frac{Y_{t}}{t}=y\right)
\end{equation}%
which characterizes deviations of the distance $Y_{t}$ from the average
value and can be expressed also trough the largest eigenvalue $\Lambda
(\gamma )$ :
\begin{eqnarray}
f(y) &=&\left( \Lambda (\gamma )-\gamma \frac{d\Lambda (\gamma )}{d\gamma }%
\right) -\gamma y \\
y &=&\frac{d}{d\gamma }\Lambda (\gamma ).
\end{eqnarray}

Thus, we have to find the dependence of $\Lambda (\gamma )$ on the parameter
$\gamma $. The master equation Eq.(\ref{masterF}) and the boundary
conditions Eq.(\ref{boundaryF}) allow one to use the Bethe ansatz in a usual
form
\begin{equation}
F_{t}(x_{1},\ldots ,x_{p})=e^{\Lambda t}\sum_{\sigma _{(1,\ldots
,p)}}A(z_{\sigma _{1}},\ldots ,z_{\sigma _{p}})~z_{\sigma
_{1}}^{-x_{1}}\ldots z_{\sigma _{p}}^{-x_{p}},  \label{ansatz}
\end{equation}
where the summation is over all permutations of $\left( \sigma _{1},\ldots
,\sigma _{p}\right) .$ The eigenvalue corresponding to eigenvalue Eq.(\ref%
{ansatz}) is
\begin{equation}
\Lambda (\gamma )=R\sum_{i=1}^{p}e^{\gamma }z_{i}+L\sum_{i=1}^{p}\frac{1}{%
e^{\gamma }z_{i}}-p.  \label{EigenValue}
\end{equation}
The parameters $z_{i}$\ satisfy the Bethe equations
\begin{equation}
z_{k}^{N}=(-1)^{p-1}\prod_{j=1}^{p}\frac{1-(1-\mu )e^{\gamma }z_{k}-\mu
e^{2\gamma }z_{j}z_{k}}{1-(1-\mu )e^{\gamma }z_{j}-\mu e^{2\gamma }z_{j}z_{k}%
}  \label{BAE}
\end{equation}
which follow from the substitution of Eq.(\ref{ansatz}) into Eq.(\ref%
{boundaryF}) and the periodic boundary conditions. The largest eigenvalue of
the master equation corresponds to the stationary state of Markov process,
so, one has to choose the solution of Eqs.(\ref{BAE}) which provides the
eigenvalue of the master equation for probability of Eq.(\ref{master1}) to
be equal to zero
\begin{equation}
\lim\limits_{\gamma \rightarrow 0}\Lambda (\gamma )=0.
\end{equation}
The Perron - Frobenius theorem ensures that this eigenvalue has no crossing
in the whole range of $\gamma $.

To solve the Bethe equations, it is convenient to transform the variables $%
z_{k}$. After the change of variables
\begin{equation}
z_{k}=\frac{1-x_{k}}{1+\mu x_{k}}e^{-\gamma }
\end{equation}
the system (\ref{BAE}) can be rewritten in the following way
\begin{equation}
e^{-\gamma N}\left( \frac{1-x_{k}}{1+\mu x_{k}}\right) ^{N}=\left( -1\right)
^{p-1}\prod_{j=1}^{p}\frac{x_{k}+\mu x_{j}}{x_{j}+\mu x_{k}}  \label{bethe}
\end{equation}
Corresponding eigenvalue has the form
\begin{equation}
\Lambda (\gamma )=\sum\limits_{k=1}^{p}\left( R\,\frac{1-x_{k}}{1+\mu x_{k}}%
+L\,\frac{1+\mu x_{k}}{1-x_{k}}\right) -p  \label{eigenvalue}
\end{equation}

\section{ The limit $\protect\mu =0$.}

In the limit $\mu \rightarrow 0,$ the ASAP becomes a particular case of the
two-parameter family of exclusion processes discussed in \cite{akk2} which,
in turn, degenerates into the $n=1$ drop-push model \cite{srb} in the
totally asymmetric case $L=0$. In this case, particles perform the partially
asymmetric random walk with rates $L$ and $R$. Going right, a particle jumps
to the closest unoccupied site at its right overtaking all adjacent
particles next to it, whereas the motion to left obeys the exclusion rule.

The simple form of the Bethe equations in this case allows one to use the
method proposed in \cite{dl} to obtain a full solution of the problem. If
one introduces the parameter
\begin{equation}
B=\left( -1\right) ^{p-1}e^{-\gamma N}\prod_{j=1}^{p}x_{j}
\end{equation}%
the solution of Eqs.(\ref{bethe}) will be given by the roots of the
polynomial equation
\begin{equation}
B\left( 1-x\right) ^{N}-x^{p}=0.
\end{equation}%
To get the largest eigenvalue, one has to choose $p$ roots approaching zero
when $\gamma \rightarrow 0$. Following \cite{dl} one can use the Cauchy
theorem to evaluate the sum over these roots integrating along the contour
enclosing all the roots in a small vicinity of zero.
\begin{equation}
\sum_{j=1}^{p}f(x_{j})=\frac{1}{2\pi i}\oint dxf(x)\frac{\frac{N}{1-x}+\frac{%
p}{x}}{1-B\frac{\left( 1-x\right) ^{N}}{x^{p}}}.  \label{Cauchy}
\end{equation}%
Inserting $R\,\left( 1-x\right) +L\,/\left( 1-x\right) -1$ instead of $f(x)$%
, we get the expression for the eigenvalue in terms of series in powers of $%
B $%
\begin{equation}
\Lambda \left( \gamma \right) =\sum_{k=1}^{\infty }B^{k}\left( -1\right)
^{kp}C_{Nk-1}^{pk-1}\left( \frac{R}{k\left( 1-\rho \right) +1/N}\ -\frac{%
L\left( 1-\rho \right) }{k-1/N}\right)  \label{seriesL}
\end{equation}%
where $\rho =p/N$ and $C_{a}^{b}=a!/(b!(a-b)!)$ is the binomial coefficient.
On the other hand, one can get the expression for $\gamma $ requiring $%
\prod_{j=1}^{p}z_{j}=1,$ which is correct for the groundstate solution .
Taking the logarithm of this product and using Eq.(\ref{Cauchy}) one gets
\begin{equation}
\gamma =\frac{1}{p}\sum_{k=1}^{\infty }\frac{B^{k}}{k}\left( -1\right)
^{kp}C_{Nk-1}^{pk-1}.  \label{seriesG}
\end{equation}%
Resolving two series of Eqs.(\ref{seriesL}), (\ref{seriesG}) and using Eqs.(%
\ref{Y1}), (\ref{Y2}) and (\ref{Y3}), we obtain the expressions for
cumulants of the total distance traveled by particles: {\small
\begin{eqnarray}
&&\lim\limits_{t\rightarrow \infty }\frac{\left\langle Y_{t}\right\rangle
_{c}}{t}=N\rho \left( \frac{R}{(1-\rho )+1/N}-\frac{L\left( 1-\rho \right) }{%
1-1/N}\right)  \label{mu0V} \\
&&\lim\limits_{t\rightarrow \infty }\frac{\left\langle
Y_{t}^{2}\right\rangle _{c}}{t}=N\frac{\rho ^{2}C_{2N-1}^{2p-1}}{\left[
C_{N-1}^{p-1}\right] ^{2}}\left( \frac{L\left( 1-\rho \right) }{\left(
1-1/N\right) \left( 2-1/N\right) }+\right.  \label{mu0Delta} \\
&&\left. \frac{R}{\left( \left( 1-\rho \right) +1/N\right) \left( 2\left(
1-\rho \right) +1/N\right) }\right)  \notag \\
&&\lim\limits_{t\rightarrow \infty }\frac{\left\langle
Y_{t}^{3}\right\rangle _{c}}{t}=N^{2}{\rho }\left[ -3\frac{\left[
C_{2N-1}^{2p-1}\right] ^{2}}{\left[ C_{N-1}^{p-1}\right] ^{4}}\left( \frac{%
L\left( 1-\rho \right) }{\left( 1-1/N\right) \left( 2-1/N\right) }+\right.
\right.  \label{mu0Y3} \\
&&\left. \frac{R}{\left( \left( 1-\rho \right) +1/N\right) \left( 2\left(
1-\rho \right) +1/N\right) }\right) +  \notag \\
&&4\frac{C_{3N-1}^{3p-1}}{\left[ C_{N-1}^{p-1}\right] ^{3}}\left( \frac{%
L\left( 1-\rho \right) }{\left( 1-1/N\right) \left( 3-1/N\right) }+\right.
\notag \\
&&\left. \left. \frac{R}{\left( \left( 1-\rho \right) +1/N\right) \left(
3\left( 1-\rho \right) +1/N\right) }\right) \right]  \notag
\end{eqnarray}%
}The scaling limit, $N\rightarrow \infty $, of these expressions is of
interest for us because it provides information about the large scale
behavior independent of the details of microscopic dynamics. While the ASEP
keeps \ the same universal behavior for any value of $\rho $, scaling
properties of the ASAP may change depending on how close to critical point
the system is. In the case $\mu =0$, the critical density corresponds to
full occupation of the lattice, $\rho _{c}=1$. In the subcritical regime
which corresponds to
\begin{equation}
\left( 1-\rho \right) \gg 1/N  \label{1-rho<<1/N}
\end{equation}%
the situation is similar to the ASEP. In the subcritical region, the
generating function, $\Lambda \left( \gamma \right) $, takes the universal
scaling form, Eq.(\ref{Generating function}) which has been already obtained
for the ASEP and claimed to be universal for all models of KPZ universality
class \cite{dl}, \cite{da}. Three model-dependent constants $%
K_{1},K_{2},K_{3}$ in Eq.(\ref{Generating function}) are
\begin{subequations}
\label{K1,K2,K3,mu=0}
\begin{eqnarray}
K_{1} &=&N\rho \left( \frac{R}{(1-\rho )}-L\left( 1-\rho \right) \right) , \\
K_{2} &=&N^{-3/2}\sqrt{\frac{\rho }{2\pi \left( 1-\rho \right) }}\left(
\frac{R}{\left( 1-\rho \right) ^{2}}+L\left( 1-\rho \right) \right) , \\
K_{3} &=&N^{3/2}\sqrt{2\pi \left( 1-\rho \right) \rho }.
\end{eqnarray}%
The average velocity of particles,
\end{subequations}
\begin{equation}
V_{\infty }=\frac{1}{p}\lim\limits_{t\rightarrow \infty }\frac{\left\langle
Y_{t}\right\rangle _{c}}{t}\simeq \frac{R}{(1-\rho )}-L\left( 1-\rho \right)
,  \label{Vinf,mu=0}
\end{equation}%
and the other cumulants of the distance travelled by particles become
divergent in the thermodynamic limit when $\rho $ approaches $1$. However,
the physical quantities characterizing the finite system should obviously be
finite. As an example, one can consider simultaneous limit $\rho \rightarrow
1,N\rightarrow \infty $. Substituting, for instance, $N^{-\vartheta
},(0<\vartheta <1)$ instead of $(1-\rho )$ into Eqs.(\ref{K1,K2,K3,mu=0},\ref%
{Vinf,mu=0}) one gets the expressions for cumulants of $Y_{t}$ which remain
finite for finite $N$. The velocity of particles in this case, $V_{\infty
}\sim N^{\vartheta }$, becomes explicitly dependent on $N$. Being finite for
finite $N$, it is divergent when $N$ tends to infinity. The upper limit for
the exponent $\vartheta $, given by Eq.(\ref{1-rho<<1/N}), is due to the
term $1/N$ \ in Eq.(\ref{seriesL}). This term plays the role of \ "infrared
cutoff" at the scale $N$, which ensures $V_{\infty }$ to remain of order $N$
if $\left( 1-\rho \right) $ becomes zero. When \ $\left( 1-\rho \right) $
becomes of order of $1/N$, the generating function, $\Lambda \left( \gamma
\right) $, looses its universal structure, Eq.(\ref{Generating function}).
Practically, this means that the presence of characteristic length $N$ \
breaks the scale invariance specific for the KPZ dynamics:%
\begin{equation*}
N\rightarrow \lambda N,\qquad K_{1}->\lambda K_{1},\qquad K_{2}->\lambda
^{-3/2}K_{2},\qquad K_{3}->\lambda ^{3/2}K_{3},
\end{equation*}%
which is held in the subcritical region, Eq.(\ref{1-rho<<1/N}). \ \ The
character of particle motion near the critical line becomes strongly
collective. Eventually, in the limit $N=p$, the process is equivalent to
totally asymmetric diffusion of a single particle, with the distance, $Y_{t}$%
, and time, $t$, rescaled as follows%
\begin{equation*}
Y_{t}\rightarrow Y_{t}N,\qquad t\rightarrow tRN.
\end{equation*}

Although the solution for $\mu=0$ allows one to approach the vicinity of
critical line, it seems to be very specific as it does not, in fact, involve
the avalanche dynamics. However, as we will see below, it catches the basic
universal scaling properties of the subcritical dynamics of the model for
arbitrary $\mu $.

\section{\protect\bigskip The case of arbitrary $\protect\mu <1$.\label{mu<1}%
}

In the case of arbitrary $\mu ,$ the Bethe equations Eq.(\ref{bethe}) cannot
be reduced to a polynomial equation. Nevertheless, they still can be solved
in the limit $N\rightarrow \infty ,p\rightarrow \infty ,\rho =p/N=const$.
Let us consider the equation obtained by taking the logarithm of both parts
of Eq.(\ref{bethe}) {\small \ }
\begin{gather}
p_{0}(x_{k})-\frac{1}{N}\sum_{j=1}^{p}\Theta \left( x_{j}/x_{k}\right)
-\gamma =2\pi iZ(x_{k})  \label{bethelog} \\
\Theta \left( y/x\right) =\ln x-\ln y+\ln \frac{1+\mu y/x}{1+\mu x/y}  \notag
\\
p_{0}=\ln \left( \frac{1-x_{k}}{1+\mu x_{k}}\right)
\end{gather}%
We define $\Theta \left( y/x\right) $ at the complex plane of variable $y$
with branch cuts shown in Fig.(\ref{Contour}). For small positive $\gamma ,$
the solution corresponding to the largest eigenvalue was shown in \cite{da}
to behave as

\begin{equation}
x_{k}\sim re^{2\pi i\frac{k}{p}},\qquad r\sim \gamma ^{1/p},\qquad \gamma
\rightarrow 0  \label{roots to zero}
\end{equation}%
The radius $r$ behaves nonanalytically$\ $when $\gamma $ approaches zero, so
in the limit $p\rightarrow \infty $ , radius $r$ becomes finite no matter
how $\gamma $ is close to zero. It can be easily verified that this is also
correct for non-zero $\mu $ at least in the limit $\gamma =0$. The
analytical function $Z(x)$ is fixed by the choice of logarithm branches. The
distribution of roots given by Eq.(\ref{roots to zero}) corresponds to the
following choice:
\begin{equation}
Z(x_{j})=-\frac{1}{N}\left( j-\frac{p+1}{2}\right) .  \label{Z(x)}
\end{equation}%
Then, assuming that the roots are at smooth contour at the complex plane,
the derivative of $Z(x)$ with minus sign has a meaning of density of roots
along the contour
\begin{equation}
R(x)=-\frac{\partial Z(x)}{\partial x}.  \label{Z(x)int}
\end{equation}%
Instead of Eq.(\ref{bethelog}), we are going to solve the equation
\begin{equation}
p_{0}(x)-\frac{1}{N}\sum_{j=1}^{p}\Theta \left( x_{j}/x\right) -\gamma =2\pi
iZ(x)  \label{bethe(x)}
\end{equation}%
together with Eqs.(\ref{Z(x)},\ref{Z(x)int}) under the assumption that the
density of roots is an analytical function of $x$. To solve these equations
one need to transform Eq.(\ref{bethe(x)}) to the integral form. This
procedure is not straightforward and depends very much on properties of the
function $Z(x)$, which should be first assumed and then can be checked a
posteriori.

A simplest method is based on the replacement of the sum by the integral
along some contour in complex plane in the thermodynamic limit \cite{syy}.
The analytic solution  then can be found for some special cases,
particularly for the case of the contour closed around zero \cite{N}. It  turns
out to be the particular  solution of our problem corresponding to single
value of $\gamma $, $\gamma =0$. A development of this idea is the expansion
method proposed in \cite{bs} for the investigation of the conical point of
ferroelectric six-vertex model, which  allows calculation of the leading
term of the first cumulant of $Y_{t}$ \cite{piph}. To obtaine   higher
cumulants one needs to calculate the finite size corrections to the
thermodynamic solution. This has been done for the ASEP \cite{LK} with the
help of the method of perturbative expansion of the Bethe equations proposed
by Kim \cite{k}. For the ASAP we use the modification of his approach, which
allows us to calculate the finite size corrections, avoiding some
assumptions made in the original method. At least in the leading orders the results do not
depend on these assumptions and we reproduce the results by Lee and Kim in
the particular case $\mu =-R/L$. To simplify the presentation we leave the
details of the solution for the Appendices (\ref{a1}-\ref{a3}), going
directly to the results.

The solution of the Bethe equations results in the generating function
obtained in the scaling limit $\gamma N^{3/2}=const$
\begin{equation}
\Lambda (\gamma )=\gamma K_{1}+K_{2}G(\gamma K_{3}).  \label{eigen}
\end{equation}%
Here $G(x)$ in a parametric form is defined by the relations
\begin{eqnarray}
G(x) &=&-\mathrm{Li}_{5/2}(-C),  \label{G(x)} \\
x &=&\mathrm{Li}_{3/2}(-C),  \label{x}
\end{eqnarray}%
and the function $\mathrm{Li}_{k}(x)$ is the polylogarithm defined by series
\begin{equation}
\mathrm{Li}_{k}(x)=\sum_{n=1}^{\infty }\frac{x^{n}}{n^{k}}
\end{equation}%
when $|x|<1$. For arbitrary negative $x$ , the integral definition can be
used:
\begin{equation}
\mathrm{Li}_{k}(x)=-\frac{1}{\Gamma (k)}\int\limits_{0}^{\infty }\frac{%
s^{k-1}ds}{1-x^{-1}e^{s}}  \label{Liintegral}
\end{equation}%
The equations, Eqs.(\ref{eigen}-\ref{x}), are nothing but Eqs.(\ref{Generating
function}-\ref{x(c)}) and $K_{1},K_{2},K_{3}$ are model dependent parameters
\begin{eqnarray}
K_{1} &=&N(1+\mu )\sum_{s=1}^{\infty }\frac{\left( L-R(-\mu )^{s-1}\right) }{%
1-\left( -\mu \right) ^{s}}\left( \frac{\rho }{\rho -1}\right) ^{s}s \\
K_{2} &=&N^{-3/2}\frac{1+\mu }{\sqrt{2\pi }}\times \\
&&\sum_{s=1}^{\infty }\frac{\left( L-R(-\mu )^{s-1}\right) }{1-\left( -\mu
\right) ^{s}}\left( \frac{\rho }{\rho -1}\right) ^{s}\frac{s^{2}(s-1+2\rho )%
}{\left( (1-\rho )\rho \right) ^{3/2}}  \notag \\
K_{3} &=&N^{3/2}\sqrt{2\pi (1-\rho )\rho }.
\end{eqnarray}

The function $G\left( x\right) $ has been already obtained for the ASEP \cite{dl} and
claimed to be universal for all models of KPZ universality class
\cite{da}. Simple form of the eigenvalue (\ref{eigen}) allows one to define
a general expression for the cumulants of the integrated particle current in
scaling limit%
\begin{eqnarray}
\lim\limits_{t\rightarrow \infty }\frac{\left\langle Y_{t}\right\rangle _{c}%
}{t} &=&K_{1}-K_{2}K_{3}=p(V_{AV}-V_{ASEP})  \label{Y1result} \\
\lim\limits_{t\rightarrow \infty }\frac{\left\langle Y_{t}^{n}\right\rangle
_{c}}{t} &=&K_{2}K_{3}^{n}G^{(n)}(0)=C_{AV}^{(n)}+C_{ASEP}^{(n)},\qquad
n\geqslant 2,
\end{eqnarray}%
where $G^{(n)}(0)$ is $n-$th derivative of the function $G(x)$ at $x=0$.
Here we divided the expressions for cumulants into two parts, which bring
different physical content. Particularly, $V_{AV}$ and $V_{ASEP}$,
\begin{eqnarray}
&&V_{AV}\simeq \frac{L\mu +R}{\left( \rho _{c}-\rho \right) ^{2}}\left[
F_{0}\left( \rho ,\rho _{c}\right) -\frac{F_{1}(\rho ,\rho _{c})}{N\left(
\rho _{c}-\rho \right) ^{2}}\right] ,  \label{Vav} \\
&&V_{ASEP}\simeq L(1+\mu )\left( 1-\rho \right) \left( 1+\frac{1}{N}\right) ,
\label{Vasep}
\end{eqnarray}%
give the contributions to the average velocity coming from the avalanche
part of the dynamics and its ASEP-like part respectively. The critical
density, $\rho _{c}$, is defined as follows
\begin{equation*}
\rho _{c}=\frac{1}{1+\mu },
\end{equation*}%
which gives the phase diagram shown in Fig. \ref{phasediag},
and the functions $F_{0}\left( \rho ,\rho _{c}\right) $ and $F_{1}(\rho
,\rho _{c})$ are nonsingular below and at the critical point, $\rho
=\rho _{c}$, so that the form of Eq.(\ref{Vav}) explicitly shows critical
singularities of the average velocity. While $V_{ASEP}$ gives the average
velocity of the ASEP when $\mu =-R/L$, $V_{AV}$ vanishes at the same time.
This is also the case for higher cumulants, which reproduce the results by
Lee and Kim in this limit
\begin{eqnarray}
\!\!\!\!\!\!\!\!\!\!\!\!\!\! &&C_{AV}^{(n)}\simeq N^{\frac{3(n-1)}{2}}\frac{%
L\mu +R}{\left( \rho _{c}-\rho \right) ^{4}}\left( 2\pi \rho \left( 1-\rho
\right) \right) ^{\frac{n-1}{2}}\rho G^{\left( n\right) }\left( 0\right)
F_{1}(\rho ,\rho _{c}),  \label{Gav} \\
\!\!\!\!\!\!\!\!\!\!\!\!\!\! &&C_{ASEP}^{(n)}\simeq L\left( 1+\mu \right) N^{%
\frac{3(n-1)}{2}}\frac{\left( 2\pi \rho \left( 1-\rho \right) \right) ^{%
\frac{n+1}{2}}}{2\pi }G^{\left( n\right) }\left( 0\right) .  \label{Gasep}
\end{eqnarray}%
Thus, when the density of particles, $\rho $, approaches its critical value,
$\rho _{c}$, the divergency of average velocity of particles is
characterized by the power law
\begin{equation}
V\sim V_{AV}\sim \left( \rho _{c}-\rho \right) ^{-\alpha }
\end{equation}%
with the critical exponent $\alpha =2$ . In \cite{piph} this exponent has
been shown to be nonuniversal with respect to the choice of different sets
of dynamical rules $\mu _{n}$. The other cumulants diverge as 4-th power of $%
\ \left( \rho -\rho _{c}\right) $. The corrections to these laws are given
in terms of the scaling variable, $N\ ^{-1}\left( \rho -\rho _{c}\right)
^{-2}$. The condition this variable is small defines the limits of
applicability of the perturbative scheme and bounds the subcritical region.%
\begin{equation*}
\left\vert \rho -\rho _{c}\right\vert \gg 1/\sqrt{N}.
\end{equation*}%
Closer to the critical line, we expect that the scaling will change as it
did in the case $\mu =0$. In addition, the case with arbitrary $\mu $ allows
one to consider the region of phase diagram above the critical line, where
the average velocity of particles will grow with $N$. This case requiring
the modification of \ finite size expansion scheme is not considered here.

To obtain the cumulants, one needs knowing only the behavior of eigenvalue
in the vicinity of the point $\gamma =0$. At the same time, for the large
deviation function, the whole range of $\gamma $ is relevant. The solution
considered above is valid for small negative $\gamma $. However, the
definition of $G\left( x\right) $, Eqs.(\ref{G(x)},\ref{x}), maintains
analyticity in the region $\left\vert C\right\vert <1$. This implies
\begin{equation}
\mathrm{Li}_{3/2}(-1)<\gamma K_{3}<\mathrm{Li}_{3/2}(1).  \label{gammarange}
\end{equation}%
Beyond this domain one has to consider the solutions with different choice
of the function $Z(x)$. Instead, we can directly use the analytical
continuation of $G(x)$ proposed in \cite{dl}. To probe the whole range of
negative $\gamma ,$ one can use the definition, Eqs.(\ref{G(x)},\ref{x}),
where the functions $\mathrm{Li}_{3/2}(C),\mathrm{Li}_{5/2}(C)$ are defined
in the integral representation Eq.(\ref{Liintegral}). For positive $\gamma $
outside of \ the domain Eq.(\ref{gammarange}), we use the following
expression
\begin{eqnarray}
G(x) &=&\frac{8}{3}\sqrt{\pi }\left[ -\ln (-C)\right] ^{3/2}-\mathrm{Li}%
_{5/2}(-C) \\
x &=&4\sqrt{\pi }\left[ -\ln (-C)\right] ^{1/2}+\mathrm{Li}_{3/2}(-C),
\end{eqnarray}%
where $0<C<-1$. Finally, using the definitions of $G(x)$ in different
domains of $\gamma $, we get the large deviation function in the scaling
limit%
\begin{equation}
f(y)=K_{3}H\left( \frac{y-K_{1}}{K_{2}K_{3}}\right)
\end{equation}%
where the universal function $H(x)$ is given by the following parametric
expression
\begin{eqnarray}
H(x) &=&G\left( \beta \right) -\beta G^{\prime }\left( \beta \right)  \\
x &=&G^{\prime }(\beta ).
\end{eqnarray}%
In \cite{dl} this function has been shown to be skew, i.e. to have different
asymptotic behavior for  its argument taking on a large negative or positive
value. This is so for the ASEP, since the speeding up and slowing down are
not equivalent due to the exclusion interaction. Specifically, it is much
easier to slow down process by stopping or  moving backward a single
particle than to speed it up by moving forward all particles simultaneously.
\ Similar qualitative interpretation of this asymmetry for the ASAP is also
possible. One can see from the explicit form of \ Eqs.(\ref{Y1result}-\ref%
{Vasep}), that the leading terms of contributions to average velocity coming
from the ASEP and avalanche parts of dynamics  have different signs, the
positive term corresponding to avalanches and the negative being for  the
ASEP drift. This is why the universal  function $G\left( x\right) $ and
subsequently $H\left( x\right) $ are different from it's standard ASEP form
in the minus sign before its arguments,  Eqs.(\ref{G(c)},\ref{x(c)})
(compare with Eqs. (20,21) from \cite{dl}). Thus, in our case the speeding
up and slowing down are interchanged comparing to the ASEP. Indeed, to
initiate an avalanche  one particle should move faster or slower then the
others to  reach eventually an occupied site. To prevent an
avalanche all particles should move simultaneously in the same direction.
The latter has much less probability then the former. However, comparing to
the ASEP the situation is even more peculiar. The solution of  the ASEP is
usually holds in the domain with definite direction of the drift, for
example $L<R$.  The opposite direction can be obtained by formal coordinate
inversion $x\rightarrow -x$ that is equivalent to  $L\leftrightarrow R$. \
The solution of the ASAP holds for all values of $L<1$ limited only by the
condition $|\mu |<1$. Thus, for small densities, $\rho $, and the rate of
the left driving, $L$, close to $1$  the situation may take place when $%
V_{ASEP}>V_{AV}$, so that the resulting average velocity will be negative.
This, however, does not affect the function $H(x)$ as its skewness is related
only to the avalanche direction rather than to the Poissonian drift. At the
same time full inversion transformation in the ASAP corresponds to
simultaneous transformations $L\leftrightarrow R$ and $\mu \rightarrow 1/\mu
$, which changes the direction of avalanches as well as that of Poissonian
drift. At the same time, if one considers the fluctuations of the particle
flow in the reference frame attached to the average flow of particles the
macroscopic fluctuations are quite similar to the ASEP up to change of  the
distance scale
\begin{eqnarray}
&&\frac{\left( Y_{t}-\left\langle Y_{t}\right\rangle \right) _{ASEP}}{\left(
Y_{t}-\left\langle Y_{t}\right\rangle \right) _{ASAP}}\rightarrow L(1+\mu
)\times   \notag \\
&&\left[ 1+\frac{\mu +R/L}{\mu }\sum_{s=1}^{\infty }\left( \frac{\rho }{\rho
-1}\right) ^{s}\frac{\left( -\mu \right) ^{s}}{1-\left( -\mu \right) ^{s}}%
\frac{s^{2}(s-1+2\rho )}{2\left( (1-\rho )\rho \right) ^{2}}\right] .
\end{eqnarray}

\section{\protect\bigskip Summary and discussion.}

To conclude, we have considered the asymmetric avalanche process on the
ring. To introduce the avalanche dynamics to the master equation for the
Poissonian process, we used the technique of the recurrent boundary
conditions. We have solved the master equation by the Bethe ansatz and
studied the solution corresponding to the stationary state. As a result, we
have calculated the cumulants of the integral particle current, exactly in
the case $\mu =0$ and in the scaling limit for general $\mu $. The large
deviation function has been obtained, which has the structure, typical for
models belonging to KPZ universality class. To calculate the finite size
corrections, we used the modification of the perturbative scheme proposed by
Kim in \cite{k}. While in the leading orders the standard and modified
approaches give the same results, it would be interesting to find out if
this is so in arbitrary orders. Our investigation is valid for the densities
below the critical point. However the model at the finite lattice can be
considered for an arbitrary density of particles. One may expect that the
scaling behavior\ of physical quantities should change above the critical
point. The question of interest is how to modify the scheme to study the
behavior of the model at the critical line and above.

\section{Acknowledgments.}

This work was supported by the National Science Council of the Republic of
China (Taiwan) through grant No. NSC 91-2112-M-001-056, Russian Foundation
for Basic Research through grant No. 99-01-00882, and Swiss National Science
Foundation through grant No. 7-SUPJ-62295.

\appendix

\section{Solution of the Bethe equations. \label{a1}}

First, we suppose that the roots of \ the Bethe equation are placed along
the closed contour \ $\Gamma $ encircling zero, and the solution still
preserves the invariance with respect to complex conjugation like the
solution corresponding to $\gamma \rightarrow 0$, Eq.(\ref{roots to zero}).
If we suppose also that the monovalued analytical function $Z^{-1}\left(
\frac{p+1}{2N}-\frac{j}{N}\right) $ can be defined everywhere at this
contour, we obtain the mapping $j$ $\rightarrow x_{j}$, which allow us to
use the Euler - Maclaurin summation formula for transformation of the sum
over the roots of Bethe equations into an integral along the segment $0<j<p$
in the plane $j$ with a correction term. The integral along the segment can
be mapped into the integral along contour $\Gamma $ in the plane $x$ (see
Appendix \ref{a2})
\begin{equation}
\sum_{j=1}^{p}f\left( x_{j}\right) \rightarrow \int\limits_{1}^{p}f\left(
Z^{-1}(j)\right) dj+f.s.c\rightarrow \oint f(x)R(x)dx+f.s.c.
\end{equation}%
where $f.s.c.$ is the correction term.

The functions of interest, e.g. the function $\Theta \left( y/x\right) $ can
be represented on $\Gamma $ as power series with additional logarithmic
terms. In this case, the only correction term appears which corresponds to a
contribution originating from \ the logarithm branch cut. As a result, the
equation Eq.(\ref{bethe(x)}) simplifies to the following form
\begin{equation}
p_{0}(x)-\oint\limits_{\Gamma }\Theta \left( y/x\right) R(y)dy-\gamma =i\pi
\rho -2\pi i\int\limits_{x_{0}}^{x}R(x)dx.  \label{betheint}
\end{equation}
The reference point $x_{0}$ is the cross point of the contour with the
positive part of the real axis, which can be defined as $x_{p}e^{-2\pi i}$.
The equation Eq.(\ref{betheint}) coincides with that obtained in \cite{piph}
in the limit of infinite lattice after replacing the sum by the integral and
neglecting the finite size corrections. It turns out that the solution of
Eq.(\ref{betheint}) gives the exact solution of Bethe equations, provided
that the inverse function $Z^{-1}\left( \frac{p+1}{2N}-\frac{j}{N}\right) $
is an analytical function in the segment $1<j<p$, what should be checked
afterwards from the solution obtained.

The only analytical solution of Eq.(\ref{betheint}) corresponds to the case $%
\gamma =0$. Taking into account the normalization
\begin{equation}
\oint R(x)dx=\rho .
\end{equation}%
we get the solution
\begin{equation}
R_{0}(x)=\frac{1}{2\pi i}\left( \frac{\rho }{x}+\frac{1}{1-x}\right) .
\label{R0}
\end{equation}%
According to Eq.(\ref{Z(x)int}) the function $Z(x)$ corresponding to this
solution, can be obtained by the integrating of the density $R(x)$:
\begin{equation*}
Z_{0}(x)=\frac{(p+1)}{2N}-\frac{1}{2\pi i}\left( \rho \ln \frac{x}{x_{0}}%
-\ln \frac{1-x}{1-x_{0}}\right)
\end{equation*}%
Using the definition of $Z(x)$, Eq.(\ref{Z(x)}), we obtain the equation for
the roots $x_{j}$ of the Bethe equations.
\begin{equation}
j=\frac{N}{2\pi i}\left( \rho \ln \frac{x_{j}}{x_{0}}-\ln \frac{1-x_{j}}{%
1-x_{0}}\right)  \label{j(x)}
\end{equation}%
Considering $j$ as a continuous parameter varying from $0$ to $p,$ Eq. \ (%
\ref{j(x)}) can be treated as an implicit definition of the contour \ $%
\Gamma $. It is easy to check that Eq.(\ref{j(x)}) has a solution
corresponding to a closed contour around zero when $x_{0}$ varies in the
interval $0\leqslant x_{0}\leqslant x_{\rho }$. The upper point $x_{\rho }$,
defined by the equation $\ x_{\rho }^{\rho }-\left( 1-x_{\rho }\right) \rho
^{\rho }(1-\rho )^{1-\rho }=0$, is a monotonous function of $\rho $ changing
in the range $0$ $\leqslant x_{\rho }$ $\leqslant 1/2$ when $\rho $ changes
from $0$ to $1$. Every value of $x_{0}$ less then $x_{\rho }$ corresponds to
the particular position of the contour $\Gamma $ passing through $x_{0}$. As
it was shown above, in the case of finite $p$ and $\gamma =0$ , all the
roots are collapsed into one point $x=0$. This situation is realized when $%
x_{0}=0$. The nonzero values of $x_{0}$ correspond to two limits $\gamma
\rightarrow 0,p\rightarrow \infty $ taken simultaneously. One can see from
the Eq.(\ref{roots to zero}) that if we put $\gamma \sim \phi ^{p},(1<\phi
<0)$, the value of $x_{0}$ takes different limits when $p\rightarrow \infty $
depending on $\phi $. At last, the case $x_{0}=x_{\rho }$ is realized when
the decay of $\gamma $ with growth of $p$ is slower than exponential and
particularly, when the limit $p\rightarrow \infty $ is taken for a fixed
$\gamma $ and then $\gamma $ is put to zero. This case is of special
interest for us because it gives the zero order of the solution \ we are
looking for. When $x_{0}>x_{\rho }$, the equation (\ref{j(x)}) has no
solutions corresponding to a closed contour encircling the zero point.

Whereas the dependence of $x_{\rho }$ on $\rho $ is defined by the
irrational equation Eq.(\ref{j(x)}), for $x_{0}=x_{\rho }$, the connection
between $\rho $ and the cross point of $\Gamma $ with the negative part of
real axes, $x_{c}$ is much simpler
\begin{equation}
x_{c}=\frac{\rho }{\rho-1 }.
\end{equation}%
Importance of this point lies in vanishing of the root density $R_{0}(x)$ at
$x_{c}$ :
\begin{equation}
R_{0}(x_{c})=0  \label{R0(xc)=0}
\end{equation}%
This fact is crucial for the Kim's perturbative scheme \cite{k},\cite{LK}.
Practically, it determines the range of applicability of Eq.(\ref{betheint}%
). It has been noted above that Eq.(\ref{betheint}) to be correct, the
analytical function $Z^{-1}\left( \frac{p+1}{2N}-\frac{j}{N}\right) $ should
be defined to map the segment $0\leqslant j<p$ to contour $\Gamma $ . It is
possible only if the first derivative $Z^{\prime }(x)=-$ $R(x)$ differs from
zero everywhere in a given region \cite{d}. According to Eq.(\ref{R0(xc)=0}%
), this is not the case at least at one point if $\gamma \neq 0$. Thus, we
cannot apply the above arguments to whole contour $\Gamma $. To overcome
this difficulty, we can separate $\Gamma $ into two parts and apply the
summation formula to intervals where the requirements of analyticity are
satisfied. It seems natural that all troubles with analyticity are
concentrated around the point of crossing the contour with the negative part
of the real axis.

Consider the solution for odd $p$, and small positive $\gamma $. Then, the
two roots closest to the negative part of the real axis, are conjugated to
each other and can be denoted by
\begin{equation}
x_{\left( p-1\right) /2}=x_{c}e^{-i\epsilon },\quad x_{\left( p+1\right)
/2}=x_{c}e^{i\epsilon },  \label{x(p+-1)/2}
\end{equation}%
where $x_{c}$ and $\epsilon $ are unknown real parameters. We exclude a
small part of the contour $\Gamma $ between $x_{\left( p-1\right) /2}$ and $%
x_{\left( p+1\right) /2}$ and assume the analyticity of $Z^{-1}\left( \frac{%
p+1}{2N}-\frac{j}{N}\right) $ at the segments $1<j<\left( p-1\right) /2$ and
$\left( p+1\right) /2<j<p$. This allows us to apply the Euler - Maclaurin
formula for these segments separately. However, points $j=\left( p+1\right)
/2$ and $j=\left( p-1\right) /2$ are located in the vicinity of a point on
the real axis where $Z^{-1}\left( \frac{p+1}{2N}-\frac{j}{N}\right) $ looses
its analyticity. Therefore, the higher the order of derivatives of $%
Z^{-1}\left( \frac{p+1}{2N}-\frac{j}{N}\right) $ at these points, the more
singular their behavior. Practically, this means that all terms of the
Euler-Maclaurin series have the same order in $N$. This is why an
alternative variant of the summation formula in the Abel-Plana form has been
applied in \cite{gs}. The Abel-Plana formula requires, however, the
analyticity of $Z^{-1}\left( \frac{p+1}{2N}-\frac{j}{N}\right) $ in the
strips $1<\mathrm{Re}j<\left( p-1\right) /2$ $\ $and $\left( p+1\right) /2<%
\mathrm{Re}j<p$, which is assumed below.

Using the Abel-Plana formula, we can transform the sum over the roots of
Bethe equations into two integrals along the segments $1<j<\left( p-1\right)
/2$ and $\left( p+1\right) /2<j<p$ \ with a finite size correction term $%
f.s.c.$ Then, these two integrals can be rewritten as the integral in the
plane $x$ along the closed contour $\Gamma $ minus the integral along the
small segment connecting the points $x_{\left( p-1\right) /2}$ and $%
x_{\left( p+1\right) /2}$.
\begin{eqnarray*}
\sum_{j=1}^{p}f\left( x_{j}\right) &\rightarrow &\int\limits_{1}^{\left(
p-1\right) /2}f\left( Z^{-1}(j)\right) dj+\int\limits_{\left( p+1\right)
/2}^{p}f\left( Z^{-1}(j)\right) dj+f.s.c \\
&&\quad \quad \quad \quad \quad \quad \downarrow \\
&&\oint f(x)R(x)dx-\int\limits_{x_{\left( p-1\right) /2}}^{x_{\left(
p+1\right) /2}}f(y)R(y)dy+f.s.c.
\end{eqnarray*}

To get the expression for the finite size correction term, we have to know
the behavior of the function $Z^{-1}\left( \xi \right) $ in the vicinity of
points $\xi =0,1/N$. Once we know $Z(x)$, the inverse function can be
constructed by inversion of its series at every point. However, the inverse
function $Z^{-1}\left( \xi \right) $ become singular at the points $\xi
=0,1/N$ in the limit $N\rightarrow \infty $. This should be taken into
account in constructing its series.

Consider the Taylor series of $Z(x)$ in points $x=x_{c}e^{\mp i\epsilon }$:

\begin{equation*}
Z(x_{c}e^{\mp i\epsilon }+t)=\sum\limits_{n=0}^{\infty }\frac{z_{n}^{\mp }}{%
n!}t^{n}
\end{equation*}
We can introduce new shifted variables $y=t-\sigma _{\mp }$ and consider the
series expansion in $y$.
\begin{equation}
Z(x_{c}e^{\mp i\epsilon }-\sigma _{\mp }+y)=z_{0}^{\mp }+\delta ^{\mp
}+\sum\limits_{n=1}^{\infty }\frac{b_{n}^{\mp }}{n!}y^{n},
\label{Z(x+sigma)}
\end{equation}
where
\begin{eqnarray*}
z_{0}^{-} &=&Z(x_{c}e^{-i\epsilon })=-\frac{1}{N};\quad
z_{0}^{+}=Z(x_{c}e^{i\epsilon })=0; \\
\delta _{\mp } &=&\sum\limits_{n=1}^{\infty }\frac{z_{n}^{\mp }}{n!}\sigma
_{\mp }^{n};\qquad b_{n}^{\mp }=\sum\limits_{k=n}^{\infty }\frac{z_{k}^{\mp
}\sigma _{\mp }^{k-n}}{(k-n)!}.
\end{eqnarray*}
Two signs $\mp $ , marking all parameters here are to remind that we
consider two expansions around points $x_{\left( p-1\right)
/2}=x_{c}e^{-i\epsilon }$ and $x_{\left( p+1\right) /2}=x_{c}e^{i\epsilon }$%
, and generally the parameters $\ z_{k}^{-},\delta ^{-},b_{n}^{-},\sigma
_{-} $ are different from $z_{k}^{+},\delta ^{+},b_{n}^{+},\sigma _{+}$.

The shift of parameters $\sigma _{\mp }$ is defined by the condition
\begin{equation}
b_{1}^{\mp }=0
\end{equation}%
that determines the shift of the expansion to the point where the density $%
R(x)$ is zero. Using the expansion (\ref{Z(x+sigma)}), we can construct the
inverse series \
\begin{eqnarray}
Z^{-1}\left( z_{0}^{\mp }+\frac{\xi }{N}\right) &=&\sum_{n=0}^{\infty
}a_{n}\left( \frac{1}{iN}\right) ^{\frac{n}{2}}\left( \pm i\sqrt{-\frac{\xi
}{i}-\frac{\delta N}{i}}\right) ^{n}  \label{z(-1)} \\
a_{0} &=&x_{c}e^{\mp i\epsilon }+\sigma ,\quad a_{1}=\sqrt{\frac{2}{b_{2}}}%
,\quad a_{2}=-\frac{b_{3}}{3b_{2}^{2}},  \notag \\
a_{3} &=&\frac{1}{18\sqrt{2}}\left( \frac{1}{b_{2}}\right) ^{\frac{7}{2}%
}\left( 5b_{3}^{2}-3b_{2}b_{4}\right) ,\ldots  \notag
\end{eqnarray}%
Here, we omitted the indices $\mp $ at variables $a,b,\sigma ,\delta $ still
implying two different expansions. To ensure the proper choice of a\ branch
of the square root in both expansions, one can check if Eq.(\ref{z(-1)}) for
$\xi =0$ is satisfied. \ Finally, after some algebra (see Appendix \ref{a3})
we rewrite the Bethe equations in the following form:

\begin{eqnarray}
R_{s} &=&\frac{\theta (-s-1)}{2\pi i}-\frac{1}{\pi }\frac{\left( -\mu
\right) ^{|s|}}{1-\left( -\mu \right) ^{|s|}}\times  \label{Rs} \\
&&\{\frac{1}{4N}\left( \tau _{+}x_{+}^{s}-\tau _{-}x_{-}^{s}\right)
+R_{s}\ln \frac{x_{+}}{x_{-}}+  \notag \\
&&\frac{1}{2i}\sum_{n\neq s}\frac{R_{n}}{s-n}\left(
x_{+}^{s-n}-x_{-}^{s-n}\right) +\frac{1}{4N}\left( \tau _{+}x_{+}^{s}-\tau
_{-}x_{-}^{s}\right) +  \notag \\
&&\frac{1}{4N}\sum_{n=1}^{\infty }\left( \frac{1}{2iN}\right) ^{\frac{n}{2}}%
\frac{\Gamma (\frac{n}{2}+1)}{\pi ^{\frac{n}{2}+1}}\times  \notag \\
&&\left[ c_{n,s}^{-}\left( \mathrm{Li}_{\frac{n}{2}+1}(-e^{-\pi \tau
_{-}})-i^{n}\mathrm{Li}_{\frac{n}{2}+1}(-e^{\pi \tau _{-}})\right) -\right.
\notag \\
&&\left. c_{n,s}^{+}\left( \mathrm{Li}_{\frac{n}{2}+1}(-e^{-\pi \tau
_{+}})-\left( -i\right) ^{n}\mathrm{Li}_{\frac{n}{2}+1}(-e^{\pi \tau
_{+}})\right) \right] \}  \notag
\end{eqnarray}%
if $s\neq 0$ and
\begin{equation}
\epsilon R_{0}=-\sum_{{s\neq 0},{s=-\infty }}^{\infty }R_{s}x_{c}^{-s}\frac{%
\sin \epsilon s}{s}+\frac{1}{2iN}  \label{Ro}
\end{equation}%
otherwise. For $\gamma $ and $\rho $, we have respectively
\begin{eqnarray}
\gamma &=&\frac{1}{2Ni}\sum_{n=1}^{\infty }\left( \frac{1}{2iN}\right) ^{%
\frac{n}{2}}\frac{\Gamma (\frac{n}{2}+1)}{\pi ^{\frac{n}{2}+1}}\times
\label{gamma} \\
&&\left[ \overline{c}_{n}^{-}\left( \mathrm{Li}_{\frac{n}{2}+1}(-e^{-\pi
\tau _{-}})-i^{n}\mathrm{Li}_{\frac{n}{2}+1}(-e^{\pi \tau _{-}})\right)
-\right.  \notag \\
&&\left. \overline{c}_{n}^{+}\left( \mathrm{Li}_{\frac{n}{2}+1}(-e^{-\pi
\tau _{+}})-\left( -i\right) ^{n}\mathrm{Li}_{\frac{n}{2}+1}(-e^{\pi \tau
_{+}})\right) \right] +  \notag \\
&&\frac{i}{2N}\left( \tau _{-}\ln x_{-}-\tau _{+}\ln x_{+}\right) +\frac{%
R_{0}}{2}\left( \ln ^{2}x_{-}-\ln ^{2}x_{+}\right) +  \notag \\
&&\sum_{n\neq 0}R_{s}\left( \frac{x_{+}^{-n}-x_{-}^{-n}}{n^{2}}+\frac{%
x_{+}^{-n}\ln x_{+}-x_{-}^{-n}\ln x_{-}}{n}\right)  \notag
\end{eqnarray}%
and
\begin{equation}
\rho =2\pi iR_{0}.  \label{rho}
\end{equation}%
Here we introduced notations $\tau _{\mp },x_{\mp },c_{n,s}^{\mp }$ which
are defined as follows
\begin{eqnarray}
\delta ^{\mp } &=&-\frac{i}{2N}\left( \mp i+\tau _{\mp }\right) ;\text{\quad
}x_{\mp }=x_{c}e^{\mp i\epsilon }+\sigma ^{\mp }  \label{delta,x} \\
\!\!\!\!\!\!\!\left[ \sum_{n=0}^{\infty }a_{n}^{\mp }x^{n}\right] ^{s}
&=&\sum_{n=0}^{\infty }c_{n,s}^{\mp }x^{n};\qquad \ln \left(
\sum_{n=0}^{\infty }a_{n}^{\mp }x^{n}\right) =\sum_{n=0}^{\infty }\overline{c%
}_{n}^{\mp }x^{n},  \label{cs,cbar}
\end{eqnarray}%
and $R_{s}$ are the expansion coefficients of Laurent series
\begin{equation}
R(x)=\sum_{s=-\infty }^{\infty }\frac{R_{s}}{x^{s+1}}.  \label{Rexp}
\end{equation}%
Using Eqs.(\ref{eigenvalue}),(\ref{Fexp}) together with Eqs.(\ref{A2Rs})-(%
\ref{A2rho}) we obtain the expression of eigenvalue in terms of $R_{s}$
\begin{equation}
\Lambda (\gamma )=2\pi iN\sum_{n=1}^{\infty }\left( -\mu \right)
^{-n}\Lambda _{n}R_{n},  \label{Lambda(Rn)}
\end{equation}%
where
\begin{equation*}
\Lambda _{n}=\left( L-R(-\mu )^{n-1}\right) (1+\mu )
\end{equation*}%
While the equations Eqs.(\ref{Rs}-\ref{rho}) look rather cumbersome, a
significant simplification takes place if the following conditions are
satisfied:
\begin{subequations}
\begin{eqnarray}
\tau _{+} &=&\tau _{-}=\tau ;  \label{tau+=tau-} \\
x_{+} &=&x_{-}=\widetilde{x};  \label{x+=x-} \\
a_{s}^{-} &=&a_{s}^{+}=a_{s}.  \label{a+=a-}
\end{eqnarray}%
Then, instead of Eqs.(\ref{Rs},\ref{gamma}), we get

\end{subequations}
\begin{eqnarray}
R_{s} &=&\frac{\theta (-s-1)}{2\pi i}-\frac{1}{2\pi i}\frac{\left( -\mu
\right) ^{|s|}}{1-\left( -\mu \right) ^{|s|}}\times  \notag \\
&&\frac{1}{N^{3/2}}\sum_{n=0}^{\infty }\left( \frac{i}{2N}\right) ^{n}\frac{%
\Gamma (n+\frac{3}{2})}{\pi ^{n+\frac{3}{2}}}\frac{c_{2n+1,s}}{\sqrt{2i}}%
\mathrm{Li}_{n+\frac{3}{2}}(-e^{\pi \tau })  \notag \\
\gamma &=&-\frac{1}{N^{3/2}}\sum_{n=0}^{\infty }\left( \frac{i}{2N}\right)
^{n}\frac{\Gamma (n+\frac{3}{2})}{\pi ^{n+\frac{3}{2}}}\frac{\overline{c}%
_{2n+1}}{\sqrt{2i}}\mathrm{Li}_{n+\frac{3}{2}}(-e^{\pi \tau }),
\end{eqnarray}
which together with Eq.(\ref{Lambda(Rn)}) reproduces the results of paper
\cite{k}. The conditions of Eqs.(\ref{tau+=tau-}) and (\ref{x+=x-}) \ are
those accepted in \cite{k} as assumptions. Equation (\ref{tau+=tau-}) is
equivalent to equality $\delta _{+}=\delta _{-}$ (see Eq.(32) in \cite{k}).
Equation (\ref{x+=x-}) \ is equivalent to the assumption that there is only
one point $\widetilde{x}$ where $Z^{\prime }(\widetilde{x})=0$ which is used
as the expansion center in \cite{k}. In our consideration, two complex
conjugated points , $x_{+}$ and $x_{-},$ are possible that merge into one
point $\widetilde{x}$ in the limit $p\rightarrow \infty $. The third
equality Eq.(\ref{a+=a-}) is a direct consequence of first two. Generally,
there are no obvious reasons for Eqs.(\ref{tau+=tau-}) and (\ref{x+=x-}) to
be satisfied .They can be checked a posteriori when the solution of Eqs.(\ref%
{Rs}) and (\ref{gamma}) is obtained. We checked them in first orders of the
perturbative solution and found that they are correct in the first
three orders which are necessary to reproduce the results of \cite{LK}.

To obtain the solution of the Eqs.(\ref{Rs}) - (\ref{rho}) which is
consistent with the exact solution in the case $\mu =0,$ one has to assume $%
\epsilon $ to behave as $\epsilon \sim N^{-1/2}$ when $N\rightarrow \infty $%
. Therefore, we assume the following expansion
\begin{equation}
\epsilon =\sum_{k=1}^{\infty }\frac{\epsilon _{i}}{N^{\frac{k}{2}}}.
\end{equation}%
The other values in the Eqs.(\ref{Rs}-\ref{rho}) can be represented as
similar expansions
\begin{equation}
R_{s}=\sum_{k=0}^{\infty }\frac{R_{s}^{(k)}}{N^{\frac{k}{2}}},\quad \rho
=\sum_{k=0}^{\infty }\frac{\rho _{i}}{N^{\frac{k}{2}}},\quad \gamma
=\sum_{k=3}^{\infty }\frac{\gamma _{i}}{N^{\frac{k}{2}}}.
\end{equation}%
Equation (\ref{R0}) is used as a zero order solution. Then Eqs.(\ref{Rs}-\ref%
{rho}) should be solved order by order in powers of $N^{-1/2}$. The scaling
dependence of $\gamma $ corresponds to $\gamma N^{3/2}=const$. The limit $%
\gamma \rightarrow 0$ corresponds to the limit $\epsilon _{1}\rightarrow 0$.
The other parameters $\epsilon _{2},\epsilon _{3},\ldots $ depend on the
way, how $\gamma $ approaches zero when $N\rightarrow \infty $. However, the
physical results do not depend on these parameters due to analyticity of
eigenvalue. Solving the Eqs.(\ref{Rs}-\ref{rho}) in first four orders, we
get the expression for eigenvalue given in Eqs.(\ref{eigen}-\ref{x})

\section{Evaluation of sums over roots of the Bethe equations.\label{a2}}

To evaluate the sum over roots of the Bethe equation, one can use \ the
asymptotic formula approximating the sum by the integrals
\begin{equation}
\sum_{i=n}^{m}f(j)=\int\limits_{n}^{m}f(j)dj+corr(f,n,m)
\label{summationform}
\end{equation}%
with the correction term given by the asymptotic Euler-Maclaurin series
\begin{equation}
corr(f,n,m)=\frac{1}{2}(f(m)+f(n))+\sum_{i=2}^{\infty }\frac{B_{i}}{i!}%
(f^{(i-1)}(m)-f^{(i-1)}(n))  \label{Euler-Maclaurin}
\end{equation}%
or by the Abel-Plana integral form
\begin{eqnarray}
&&corr(f,n,m)=\frac{1}{2}(f(m)+f(n))+  \label{abel-planfsc} \\
&&\frac{1}{i}\int\limits_{0}^{\infty }\frac{f(m+it)-f(m-it)-f(n+it)+f(n-it)}{%
e^{2\pi t}-1}dt.
\end{eqnarray}%
The former requires the analyticity of the function $f(j)$ at the segment of
real axes $j\in \lbrack n,m]$ and the latter at the strip of complex plane, $%
\mathrm{Re}x\in \lbrack n,m].$ Let us suppose, that the analytical structure
of the function $Z(x)$ allows one to define the analytical inverse function $%
Z^{-1}\left( \frac{p+1}{2N}-\frac{j}{N}\right) $ that maps the segment $j\in
\lbrack 0,p]$ into the closed contour $\Gamma $ encircling zero in the plane
of the variable $x$. Then the derivatives with respect to $j$ can be
expressed in terms of $x_{j}$ as follows
\begin{equation}
\frac{\partial }{\partial j}f\left( j\right) \rightarrow -\frac{1}{NR(x_{j})}%
f^{\prime }\left( -Z\left( x_{j}\right) N+\frac{p+1}{2}\right) .
\label{df(j)/dj}
\end{equation}%
We are interested in calculation of sums of the form
\begin{equation}
\frac{1}{N}\sum_{j=1}^{p}F(x_{j}),  \label{sumf(x_j)}
\end{equation}%
where $F(x)$ can be represented at the contour $\Gamma $ as Laurent series
with additional logarithmic term
\begin{equation}
F(x)=\overline{F}\ln x+\sum_{s=-\infty }^{\infty }F_{s}x^{s}.  \label{Fexp}
\end{equation}%
The root $x_{p}$ lies at the real part of positive axes. We can introduce
its \textquotedblright twin\textquotedblright\ at the other side of
logarithm branch cut, $x_{0}=e^{-2\pi i}x_{p}$, which corresponds to $j=0$.
Then, application of Euler-Maclaurin formula Eq.(\ref{Euler-Maclaurin})
gives
\begin{equation}
\frac{1}{N}\sum\limits_{j=1}^{p}F(x_{j})=\frac{1}{N}\left(
\sum\limits_{j=0}^{p}F(x_{j})-F(x_{0})\right) =\oint\limits_{\Gamma
}F(y)R(y)dy+\frac{\pi i}{N}\overline{F}  \label{sumtoint0}
\end{equation}%
Indeed, all derivatives in the Euler-Maclaurin series Eq.(\ref%
{Euler-Maclaurin}) taken at the points $x_{0}$ and $x_{p}$, being equal,.are
cancelled by each other. The only contribution to correction term comes from
difference between imaginary parts of the logarithm at the banks of its
branch cut. As it is shown in the section (\ref{mu<1}), this case is limited
by $\gamma =0$.

Let us consider the case when the roots $x_{\left( p-1\right) /2}$ and $%
x_{\left( p+1\right) /1}$ are located in the vicinity of the point $x_{c}$
satisfying Eq.(\ref{R0(xc)=0}), and, therefore, we can not guarantee
existence of an analytical function $Z^{-1}\left( \frac{p+1}{2N}-\frac{j}{N}%
\right) $ that maps the segment $j\in \lbrack 0,p]$ to closed contour $%
\Gamma $. We, however, still assume, that the mapping like this exists at
two its segments, which connect points $x_{0},x_{\left( p-1\right) /2}$ and $%
x_{\left( p+1\right) /1},x_{p}$. One can see from Eq.(\ref{df(j)/dj}), that
every derivative with respect to $j$ brings the competitive coefficients $%
1/N $ and $1/R(x_{j})$, i.e. when $N$ tends to infinity $R\left( x_{\left(
p\pm 1\right) /1}\right) $ goes to zero. This is why we use Abel-Plana
summation formula Eq.(\ref{abel-planfsc}) instead of Eq.(\ref%
{Euler-Maclaurin}) to take into account all contributions of the same order
in $N$. Applying it to each of two segments separately and using formula Eq.(%
\ref{x(p+-1)/2}) for the roots $x_{\left( p-1\right) /2}$ and $x_{\left(
p+1\right) /1}$ we obtain
\begin{eqnarray}
\frac{1}{N}\sum\limits_{j=1}^{p}F(x_{j}) &=&\sum_{s=-\infty }^{\infty
}R_{s}(I_{s}^{0}+I_{s}^{\epsilon })+\frac{\pi i}{N}\overline{F}+
\label{sumtoint} \\
&&\overline{F}\left( \frac{\ln x_{c}}{N}+\overline{\varkappa }\right)
+\sum_{s=-\infty }^{\infty }F_{s}\left( \frac{x_{c}^{s}\cos \epsilon s}{N}+%
\frac{\varkappa _{s}}{N}\right)  \notag
\end{eqnarray}

where
\begin{eqnarray}
I_{s}^{0} &=&2\pi i\left\{
\begin{array}{cc}
F_{s}-\frac{\overline{F}}{sx_{0}^{s}} & s\neq 0 \\
F_{0}+\overline{F}\left( \ln x_{0}+i\pi \right) & s=0%
\end{array}
\right. ,  \label{I0} \\
I_{s}^{\epsilon } &=&-2i\epsilon \left\{
\begin{array}{cc}
\begin{array}{c}
F_{s}+\sum\limits_{{n\neq s},{n=-\infty }}^{\infty }F_{n}\frac{%
x_{c}^{n-s}\sin \left( n-s\right) \epsilon }{\epsilon \left( n-s\right) }+
\\
\frac{\overline{F}}{sx_{c}^{s}}\left( \left( \ln x_{c}+\frac{1}{s}\right)
\frac{\sin \epsilon s}{\epsilon }-\cos \epsilon s\right)%
\end{array}
& s\neq 0 \\
F_{0}+\sum\limits_{{n\neq 0},{n=-\infty }}^{\infty }F_{n}\frac{x_{c}^{n}\sin
n\epsilon }{\epsilon n}+\overline{F}\ln x_{c} & s=0%
\end{array}
\right. ,
\end{eqnarray}
\begin{eqnarray}
\varkappa _{s} &=&\frac{1}{i}\int\limits_{0}^{\infty }\left\{ \left[ Z^{-1}(-%
\frac{1+it}{N})\right] ^{s}-\left[ Z^{-1}(-\frac{1-it}{N})\right] ^{s}\right.
\label{kappas} \\
&&\left. -\left[ Z^{-1}(-\frac{it}{N})\right] ^{s}+\left[ Z^{-1}(\frac{it}{N}%
)\right] ^{s}\right\} /\left( e^{2\pi t}-1\right) dt,  \notag \\
\overline{\varkappa } &=&\frac{1}{i}\int\limits_{0}^{\infty }\left\{ \ln
Z^{-1}(-\frac{1+it}{N})-\ln Z^{-1}(-\frac{1-it}{N})\right.  \label{kappabar}
\\
&&\left. -\ln Z^{-1}(-\frac{it}{N})+\ln Z^{-1}(\frac{it}{N})\right\} /\left(
e^{2\pi t}-1\right) dt  \notag
\end{eqnarray}
and $R_{s}$ are the coefficients of the Laurent expansion of the density
defined in Eq.(\ref{Rexp}).

\section{ Derivation of equations for $R_{s}$.\label{a3}}

Rewriting the sum in the Eq.(\ref{bethe(x)}) with the help of Eq.(\ref%
{sumtoint}) and collecting coefficient of the same powers of $x$ we get
\begin{eqnarray}
s &\neq &0:  \notag \\
&&R_{s}=\frac{\theta (-s-1)}{2\pi i}-\frac{1}{\pi }\frac{\left( -\mu \right)
^{|s|}}{1-\left( -\mu \right) ^{|s|}}\times  \label{A2Rs} \\
&&\left( \epsilon R_{s}+\sum_{{n\neq s},{n=-\infty }}^{\infty }\frac{%
x_{c}^{s-n}}{s-n}\sin \epsilon \left( s-n\right) R_{n}-\frac{x_{c}^{s}}{2iN}%
\cos \epsilon s-\frac{\varkappa _{s}}{2iN}\right)  \notag \\
s &=&0:  \notag \\
&&\epsilon R_{0}=-\sum_{{s\neq 0},{s=-\infty }}^{\infty }R_{s}x_{c}^{-s}%
\frac{\sin \epsilon s}{s}+\frac{1}{2iN}  \label{A2R0} \\
&&\gamma =2i\sum_{{s\neq 0},{s=-\infty }}^{\infty }\frac{x_{c}^{-s}}{s}%
\left( \epsilon \cos \epsilon s-\frac{\sin \epsilon s}{s}\right) R_{s}+\frac{%
\overline{\varkappa }}{N}  \label{A2gamma} \\
&&\rho =2\pi iR_{0}  \label{A2rho}
\end{eqnarray}
where we use the function $\Theta (y/x)$, treated as a function of the
variable $y,$ as the expansion valid at the contour $\Gamma $ with
coefficients defined like in Eq.(\ref{Fexp})
\begin{equation*}
\Theta _{n}=\left\{
\begin{array}{cc}
\frac{\left( -\mu \right) ^{|s|}}{sx^{s}} & s\neq 0 \\
\ln x & s=0%
\end{array}
\right. ;\qquad \overline{\Theta }=-1.
\end{equation*}
To go further one needs to obtain the explicit expressions for $\varkappa
_{s},\overline{\varkappa }$ in terms of $R_{s}$. Using the expansion of $%
Z^{-1}\left( z_{0}^{\mp }+\frac{\xi }{N}\right) $, Eq.(\ref{z(-1)}), we get
\begin{eqnarray}
\varkappa _{s} &=&\frac{1}{2}\sum_{n=1}^{\infty }\left( \frac{1}{2iN}\right)
^{\frac{n}{2}}\left[ c_{n,s}^{-}Y_{n}^{-}-c_{n,s}^{+}Y_{n}^{+}\right]
\label{kappas(Y)} \\
\overline{\varkappa } &=&\frac{1}{2}\sum_{n=1}^{\infty }\left( \frac{1}{2iN}%
\right) ^{\frac{n}{2}}\left[ Y_{n}^{-}\overline{c}_{n}^{-}-Y_{n}^{+}%
\overline{c}_{n}^{+}\right] ,  \label{kappa0(Y)}
\end{eqnarray}
where the coefficients $\overline{c}_{n}^{\mp }$ and $c_{n,s}^{\mp }$ are
defined in Eq.(\ref{cs,cbar}) an the functions $Y_{n}^{\mp }$ are given by
\begin{equation}
Y_{n}^{\mp }=\frac{1}{i}\int\limits_{0}^{\infty }\frac{\left[ \left( \pm i%
\sqrt{-t+\tau _{\mp }-i}\right) ^{n}-\left( \pm i\sqrt{t+\tau _{\mp }-i}%
\right) ^{n}\right] }{e^{\pi t}-1}dt
\end{equation}
The method of evaluation of these integrals is described in detail in \cite%
{LK}. Finally, for $Y_{n}^{\mp }$ we get
\begin{eqnarray}
Y_{n}^{\mp } &=&\frac{1}{i}\left\{ \left( \mp i-\frac{\tau _{\mp }\mp i}{%
\frac{n}{2}+1}\right) \left( \pm i\sqrt{\tau _{\mp }\mp i}\right)
^{n}+\right. \\
&&\left. \frac{\Gamma (\frac{n}{2}+1)}{\pi ^{\frac{n}{2}+1}}\left( \mathrm{Li%
}_{\frac{n}{2}+1}(-e^{-\pi \tau _{\mp }})-\left( \pm i\right) ^{n}\mathrm{Li}%
_{\frac{n}{2}+1}(-e^{\pi \tau _{\mp }})\right) \right\} .  \notag
\end{eqnarray}
Using the equalities
\begin{equation}
\sum_{n=1}^{\infty }c_{n,s}^{\mp }\left( \pm i\sqrt{\tau _{\mp }-i}\right)
^{n}\left( \frac{1}{2iN}\right) ^{\frac{n}{2}}=\left[ Z^{-1}(z_{0}^{\mp })%
\right] ^{s}-c_{0,s}^{\mp },
\end{equation}
\begin{equation}
\sum_{n=1}^{\infty }\overline{c}_{n}^{\mp }\left( \pm i\sqrt{\tau _{\mp }-i}%
\right) ^{n}\left( \frac{1}{2iN}\right) ^{\frac{n}{2}}=\ln \left[
Z^{-1}(z_{0}^{\mp })\right] -\overline{c}_{0}^{\mp },
\end{equation}
\begin{eqnarray}
&&\sum_{n=1}^{\infty }c_{n,s}^{\mp }\left( \pm i\sqrt{\tau _{\mp }-i}\right)
^{n}\left( \frac{1}{2iN}\right) ^{\frac{n}{2}}\frac{\tau _{\mp }\mp i}{\frac{%
n}{2}+1}=\delta ^{\mp }Nc_{0,s}^{\mp }+ \\
&&\qquad \qquad \qquad \qquad \qquad \qquad \qquad \qquad
N\int\limits_{c_{0,s}^{\mp }}^{Z^{-1}\left( z_{0}^{\mp }\right) }x^{s}R(x)dx,
\notag
\end{eqnarray}
\begin{eqnarray}
&&\sum_{n=1}^{\infty }\overline{c}_{n}^{\mp }\left( \pm i\sqrt{\tau _{\mp }-i%
}\right) ^{n}\left( \frac{1}{2iN}\right) ^{\frac{n}{2}}\frac{\tau _{\mp }\mp
i}{\frac{n}{2}+1}=N\delta ^{\mp }\overline{c}_{0}^{\mp }+ \\
&&\qquad \qquad \qquad \qquad \qquad \qquad \qquad
N\int\limits_{c_{0,s}^{\mp }}^{Z^{-1}\left( z_{0}^{\mp }\right) }\ln xR(x)dx,
\notag
\end{eqnarray}
\begin{equation}
Z^{-1}(z_{0}^{\mp })=x_{c}e^{\mp i\epsilon };\qquad c_{0,s}^{\mp }=x_{\mp
}^{s};
\end{equation}
and Eqs.(\ref{delta,x}) we come to the system of equations for $R_{s}$, Eqs.(%
\ref{Rs})-(\ref{gamma}).

\newpage

\begin{figure}[h]
\caption{The asymmetric avalanche process.}
\label{ASAP}
\end{figure}

\begin{figure}[h]
\caption{The analytical structure of the contour $\Gamma $.
Zigzag lines show the branch cuts of the function $\Theta (y/x)$ in the
complex plane of the variable $y$. The broken segment of the contour should
be excluded from integration when $\protect\gamma \neq 0$.}
\label{Contour}
\end{figure}

\begin{figure}[h]
\caption{The phase diagram of the asymmetric avalanche
process.}
\label{phasediag}
\end{figure}
\begin{center}

\thispagestyle{empty}
\epsfig{file=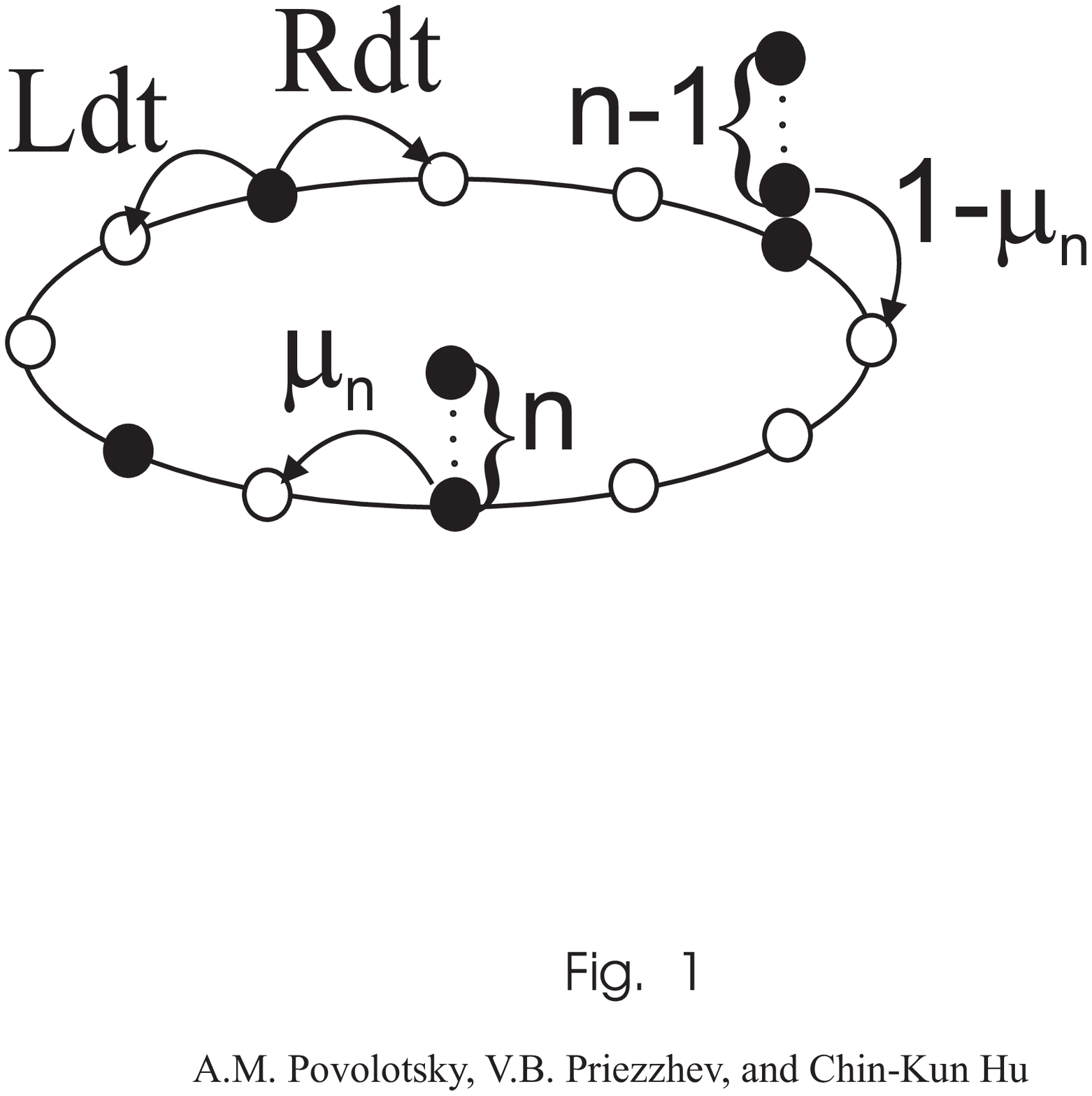,width=16cm}

\newpage
\thispagestyle{empty}
\epsfig{file=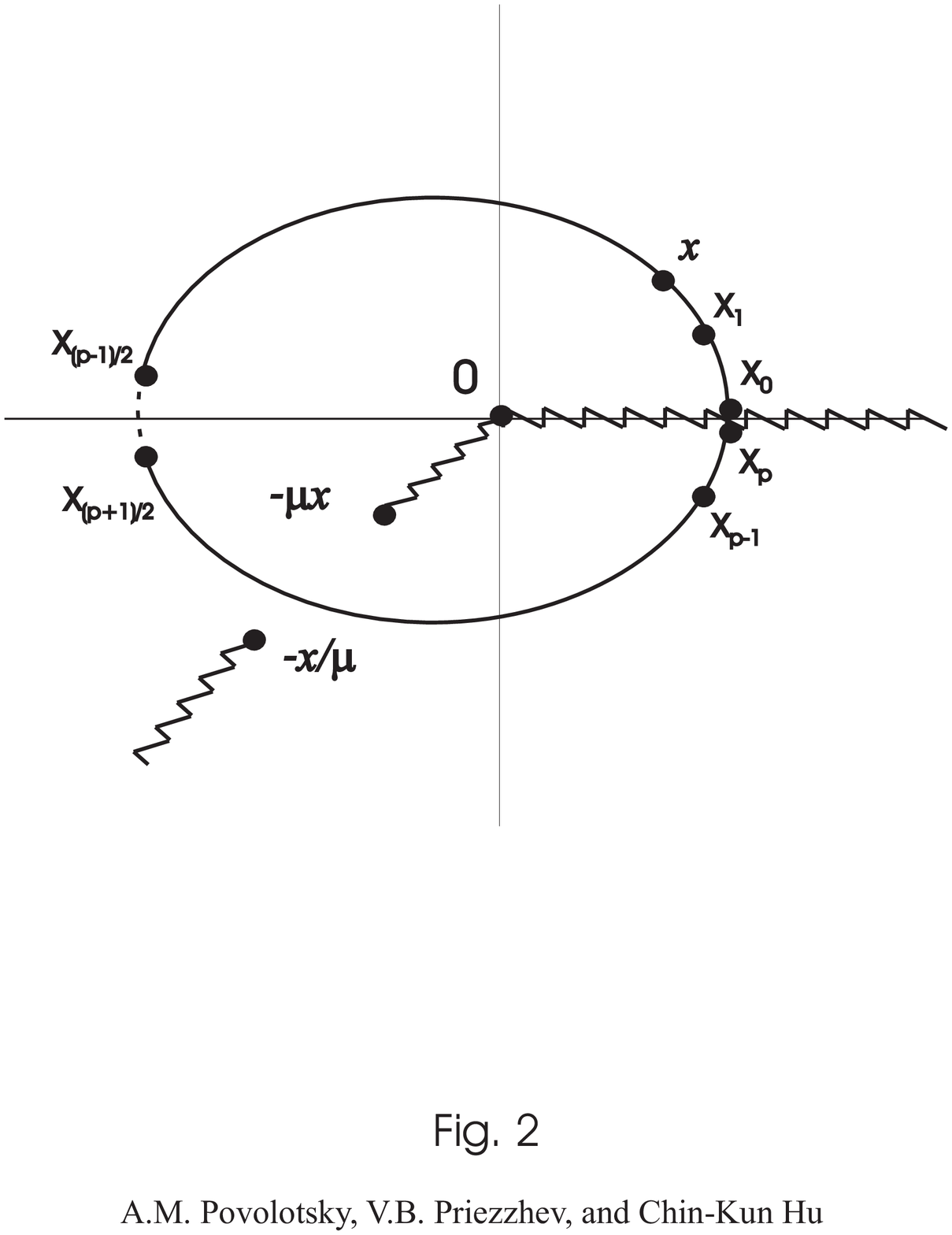,width=16cm}

\newpage
\thispagestyle{empty}
\epsfig{file=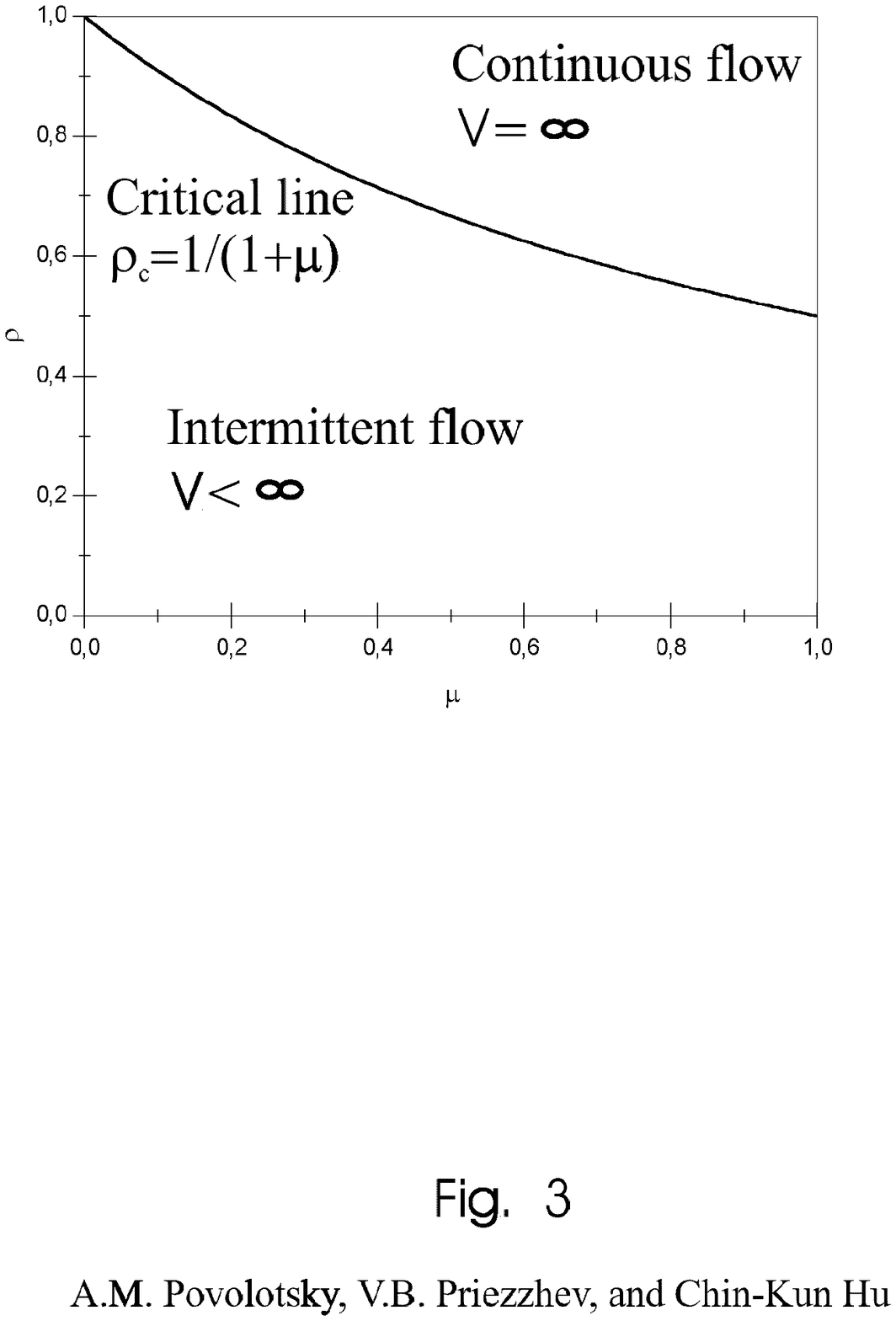,width=16cm}
\end{center}

\end{document}